\documentstyle[12pt]{article}

\newcommand{\nsect}{\setcounter{equation}{0}
\def\theequation{\thesection.\arabic{equation}}\section}
\def\simlt{\mathrel{\lower2.5pt\vbox{\lineskip=0pt\baselineskip=0pt
           \hbox{$<$}\hbox{$\sim$}}}}
\def\simgt{\mathrel{\lower2.5pt\vbox{\lineskip=0pt\baselineskip=0pt
           \hbox{$>$}\hbox{$\sim$}}}}
\def\pr  {{\sl Phys. Rev.}~{\bf C}}

\def\np  {{\sl Nucl. Phys.}~{\bf B}}

\def\pl  {{\sl Phys. Lett.}~{\bf B}}

\def\MPL{{\it Mod. Phys. Lett. }}
\def\SPJ{{\it Sov. J. Phys. }}
\def\PR{{\it Phys. Rev. }}
\def\AM{{\it Adv. Math. }}
\def\IJMP{{\it Int. J. Mod. Phys. }}
\newcommand{\bea}{\begin{eqnarray}}
\newcommand{\eea}{\end{eqnarray}}
\newcommand{\bean}{\begin{eqnarray*}}
\newcommand{\eean}{\end{eqnarray*}}
\begin{document}
\begin{titlepage}
\vspace*{-1cm}
\hfill{CERN-TH.7148/94}\\
\phantom{bla}
\hfill{hep-th/9402073}\\
\phantom{bla}
\hfill{LPTENS 94/04}\\
\phantom{bla}
\hfill{CPTH-A291.0294}\\
\phantom{bla}
\hfill{UCLA/94/TEP/5}\\
\vskip 0.3cm
\normalsize
\begin{center}
{\Large\bf Exact Supersymmetric String Solutions in Curved
Gravitational Backgrounds}
\end{center}
\vskip .3cm
\begin{center}
{\bf I. Antoniadis}
\\Centre de Physique Th\'eorique,
Ecole Polytechnique, 91128 Palaiseau, France\footnote{\it Laboratoire
Propre du CNRS UPR A.0014}\\
\vskip .2cm
{\bf S. Ferrara} and {\bf C. Kounnas}\footnote{On leave from Ecole
Normale
Sup\'erieure, 24 rue Lhomond, 75231 Paris Cedex 05, France.}
\\Theoretical Physics Division, CERN, Geneva, Switzerland\\
\end{center}

\begin{abstract}
\noindent
We construct a new class of exact and stable superstring solutions
based on $N=4$ superconformal world-sheet symmetry. In a subclass of
these, the full spectrum of string excitations is derived in a
modular-invariant way.

In the weak curvature limit, our solutions describe a target
space with non-trivial metric and topology, and generalize the
previously known (semi) wormhole. The effective field
theory limit is identified in certain cases, with
solutions of the $N=4$ and $N=8$ extended gauged supergravities, in
which the number of space-time supersymmetries is reduced by a factor
of 2 because of the presence of non-trivial dilaton, gravitational
and/or gauge backgrounds.

In the context of string theory, our solutions correspond to stable
non-critical superstrings in the strong coupling region; the
super-Liouville field couples to a unitary matter system with central
charge $5\le{\hat c}_M\le 9$.
\end{abstract}
\vspace*{.3cm}
\begin{flushleft} CERN-TH.7148/94 \\
February 1994
\end{flushleft}
\vfill\eject
\end{titlepage}
%


\nsect{Introduction}

The study of string propagation in non-trivial gravitational
backgrounds can provide a better understanding of quantum
gravitational phenomena at short distances. Non-trivial classical
string backgrounds can be obtained by two
different methods. The first makes use of a two-dimensional
$\sigma$-model where the space-time backgrounds correspond to
field-dependent coupling constants \cite{betafunct}.
The vanishing of the corresponding
$\beta$-functions is identified with the background field equations
of motion in the target space \cite{betafunct}-\cite{KKLn24}.
The second approach consists of replacing the free space-time
coordinates by a non-trivial (super)conformal system, which, in the
semiclassical limit, can be interpreted as describing a string
propagation in non-trivial
space-time \cite{ABEN}-\cite{nwgrwaves}.

The two methods are useful and complementary. The $\sigma$-model
approach
provides a clear geometric interpretation, but it has the
disadvantage of
the $\alpha^{\prime}$-expansion, which is valid only when all
curvatures and
derivatives on space-time fields are small. In this way, one can
easily obtain approximate solutions, but their possible extension to
exact ones is in general difficult to prove. The conformal field
theory approach takes into account all orders in $\alpha^{\prime}$
automatically and has the main advantage of deriving exact string
vacua. However, the background interpretation of a given
exact string solution is, in general, an ill-defined notion
\cite{n4kounnas}.
Indeed, the notion of  space-time dimensionality and topology breaks
down for
a solution that involves highly curved backgrounds, namely when the
metric and/or gauge field curvatures are of the order of the string
scale.

A typical example concerning the dimension and topology of space-time
is that of the $SU(2)$ level $k$ group manifold
compactification. For
large $k$
(small curvature) the target space is a three-dimensional sphere
$S^3$. For small $k$ (high
curvature) this background interpretation fails. It is in fact well
known that the $SU(2)_{k=1}$ WZW  model is equivalent to a $c = 1$
conformal system
defined by one free bosonic co-ordinate compactified on a cycle of
radius $R = \frac{2}{R} = \sqrt{2}$ (self-dual point).
Naively one may interpret this toroidal compactification as
a one-dimensional space with $S^1$ topology, which is in
contradiction
with the three-dimensional interpretation with $S^3$ topology of
$SU(2)_{k=1}$. This shows that both the dimensionality and the
topology of target space are not well-defined concepts in string
theory. In general, a background interpretation of a given string
solution
exists only when the lower Kaluza-Klein excitations have masses much
smaller than the typical string scale $(M_{st} = \alpha '^{-1/2})$.

In this work we present a special class of exact solutions of the
heterotic and type-II superstrings, which are based on some $N=4$
superconformal systems \cite{n4kounnas}. According to the realization
of the
underlying superconformal algebra, our solutions are classified into
different subclasses.   More explicitly, we arrange the degrees of
freedom of the ten
supercoordinates in three superconformal systems:
\begin{equation}
\{ \hat c\} = 10 = \{ \hat c = 2 \} + \{ \hat c = 4 \}_1 + \{ \hat c
= 4\}_2~.
\label{ceq}
\end{equation}
The $\hat c = 2$ system is saturated by two free superfields. In one
variation of our solutions, one of the two free superfields is chosen
to be the time-like supercoordinate and the other to be one of the
nine space-like supercoordinates. In other variations, both
supercoordinates are Euclidean or even compactified on a one- or
two-dimensional torus. The remaining eight supercoordinates appear
in  groups of four in $\{\hat c = 4\}_1$ and $\{\hat c = 4\}_2$. Both
$\{\hat c = 4\}_A$ systems exhibit $N = 4$ superconformal symmetry of
the Ademollo et al. type \cite{ademolo}. The non-triviality of our
solutions follows from the fact that there exist realizations of the
$\hat c = 4$, $N = 4$ superconformal systems that are based on
geometrical and topological non-trivial spaces, other than the
$T^4/Z_2$ orbifold and the $K_3$ compact manifold \cite{n4kounnas}.

The first subclass is characterized by two integer parameters $k_1$,
$k_2$, which are the levels of two $SU(2)$ group manifolds. For
weakly
curved backgrounds (large $k_A$) these solutions can be interpreted
in terms of a ten-dimensional, but topologically non-trivial, target
space of the form $R^4 \otimes S^3 \otimes S^3$. In the special limit
$k_2\rightarrow \infty$ one obtains the semi-wormhole solution of
Callan, Harvey and Strominger \cite{wormclas}, based on a
six-dimensional flat
background, combined with a four-dimensional space
$W_{k_1}^{(4)} \equiv U(1)\otimes SU(2)_{k_1}$, which describes the
semi-wormhole. The underlying superconformal field theory associated
to $W_{k_1}^{(4)}$ includes a supersymmetric $SU(2)_{k_1}$ WZW model
describing the three coordinates of $S^3$ as well as a non-compact
dimension with a background charge, describing the scale factor of
the sphere. Furthermore, it was known that the {\it five-brane}
background \cite{5brane} $M^6\otimes W_{k_1}^{(4)}$  admits two
covariantly constant
spinors and, therefore, respects up to two space-time supersymmetries
consistently with the $N=4$ symmetry of the $W_{k_1}^{(4)}$
superconformal system \cite {wormclas}.
The explicit representation of the desired $N=4$ algebra is derived
in
\cite{kprn4},\cite{n4kounnas}, while the target space interpretation
as a four-dimensional semi-wormhole space is given in
\cite{wormclas}. In the
context of this interpretation, the 10-d backgrounds
of the first subclass of our solutions is that of a product of
topologically non-trivial spaces, $M^{2}\otimes W_{k_1}^{(4)} \otimes
W_{k_2}^{(4)}$ ($M^2$ is the flat (1+1) space-time).

A second subclass of solutions is based on a different realization of
the $N=4$ superconformal system with $\hat c=4$
\cite{n4kounnas},\cite{KKLn24}. Here one
replaces the $W_k^{(4)}$ space by a new $N=4$ system,
$\Delta_k^{(4)} \equiv \bigg\{\bigg [\frac{SU(2)}{U(1)}\bigg]_k
\otimes \bigg[ \frac{SL(2,R)}{U(1)}\bigg]_{k+4}\bigg\}_{\rm SUSY}$,
i.e. a gauged supersymmetric WZW model with $\hat c[\Delta^{(4)}_k] =
4$ for any  value of $k$. The choice of the levels $k$ and $k+4$ is
necessary for the existence of an $N=4$ symmetry with $\hat c=4$.
Using $\Delta_k^{(4)}$ or $W_k^{(4)}$ as four-dimensional subspaces,
we can construct non-trivial 10-d solutions, which admit $N=2$ target
space supersymmetries in the heterotic string, or $N=2+2$ target
space supersymmetries in type-II strings.

Another subclass of solutions is obtained using the dual space of
$W_k^{(4)}$, $C^{(4)}_k$ \cite{RoVe},\cite{n4kounnas},\cite{KKLn24}.
It turns out that the $C^{(4)}_k$ conformal
system with $\hat c=4$ shares with  $\Delta_k^{(4)}$ and $W_k^{(4)}$
the same $N=4$ superconformal properties. The explicit
realization of the $C^{(4)}_k$ space is given in \cite{n4kounnas}.
{}From the conformal theory viewpoint $C^{(4)}_{k}$  is based on the
supersymmetric gauged WZW model
$C^{(4)}_k \equiv \bigg(\frac{SU(2)}{U(1)}\bigg)_{k}\otimes
U(1)_{R}\otimes U(1)_{Q}$ with a background charge
$Q=\sqrt{\frac{2}{k+2}}$ in  one of the two coordinate currents
($U(1)_{Q}$). The other free  coordinate ($U(1)_{R}$) is compactified
on a torus with radius $R=\sqrt{2k}$.

Having at our disposal non-trivial $N=4$, ${\hat c}=4$
superconformal systems, we can use them as building blocks to obtain
new classes of $exact$ and $stable$ string solutions around
non-trivial backgrounds in both type-II and heterotic superstrings,
as
shown in (\ref{ceq}). Some typical 10-d target spaces, obtained via
the above-mentioned conformal block construction, are:
\begin{eqnarray}
{\rm A}) &i) F^{(2)}\otimes W_{k_1}^{(4)} \otimes W_{k_2}^{(4)}
\nonumber\\
&ii) F^{(2)}\otimes F^{(4)} \otimes W_{k}^{(4)}
\nonumber\\
{\rm B}) &i) F^{(2)}\otimes C_{k_1}^{(4)} \otimes C_{k_2}^{(4)}
\nonumber\\
&ii) F^{(2)}\otimes F^{(4)} \otimes C_{k}^{(4)}
\nonumber\\
{\rm C}) &i) F^{(2)}\otimes C_{k_1}^{(4)} \otimes W_{k_2}^{(4)}
\nonumber\\
&ii) F^{(2)}\otimes C_{k_1}^{(4)} \otimes \Delta_{k_2}^{(4)}
\nonumber\\
&iii) F^{(2)}\otimes \Delta_{k_1}^{(4)} \otimes W_{k_2}^{(4)}
\nonumber\\
{\rm D}) &i) F^{(2)}\otimes \Delta_{k_1}^{(4)}
\otimes\Delta_{k_2}^{(4)}
\nonumber\\
&ii) F^{(2)}\otimes F^{(4)} \otimes \Delta_k^{(4)}
\label{constr}
\end{eqnarray}
In the above expressions, $F^{(4)}$ stands for a four-dimensional
Lorentzian flat space, compact or non-compact, as well as for a
four-dimensional $T^4/Z_2$ orbifold; $F^{(2)}$ denotes also a
two-dimensional flat space, compact or non-compact, with Lorentzian
or
Euclidean signature. Note that the Euclidean version of the three
cases (ii) (i.e. when $F^{(2)}\otimes F^{(4)}$ is a compact
six-dimensional flat space) can be identified with three different
kinds of four-dimensional gravitational and/or dilatonic instanton
solutions. In this interpretation, the $W_k^{(4)}$, $C_k^{(4)}$, or
${\Delta}_k^{(4)}$ subspace describes the Euclidean version of our
space-time.

In Section 2, we discuss the connection of our constructions
with stable solutions of extended gauge supergravities (type-A)
and their relation with non-critical superstrings (type-A and -C).
In Section 3, we describe the four different realizations of the
$N=4$ ${\hat c}=4$ superconformal algebra, used to define the
conformal blocks appearing in (\ref{constr}). In Section 4, we
derive the spectrum of string excitations and give examples of
one-loop partition functions for the semi-wormhole space (A-ii), as
well
as for the general (A-i) solution. In particular, we discuss the
special values $k_1=2$, $k_2=\infty$ and $k_1=k_2=0$ corresponding
to non-critical superstrings with ${\hat c}_M=8$ and ${\hat c}_M=5$,
respectively, where massless states appear in the twisted sector of
the theory. Finally, Section 5 contains our conclusions.

\section{Connection to gauged supergravities
 and non-critical strings}
\subsection{Gauged supergravity}
\setcounter{equation}{0}
The type-A constructions based on $W^{(4)}$ conformal blocks
describe, from the target space point of view, stable solutions of
4-d gauged supergravities \cite{CSF},\cite{Roo}, which leave
some of the
space-time supersymmetries unbroken. In fact, consider the 10-d
heterotic or
type-II superstring compactified on a product of two
three-dimensional spheres. The corresponding superconformal field
theory is then given by a supersymmetric WZW model based on a ${\bf
K}^{(6)}\equiv SU(2)_{k_1}\otimes SU(2)_{k_2}$ group manifold, where
the Kac-Moody levels $k_A$ define the radii of the spheres $r_A$,
$k_A =r_A^2$ \cite{ABS}.
In contrast with the toroidal compactification
($T^6\equiv U(1)^6$), where the six graviphotons are Abelian, in
${\bf K}^{(6)}$ compactification they become non-Abelian. As expected
from field theory Kaluza-Klein arguments, in the large radii limit
the resulting effective theory is an $SU(2)_{k_1}\otimes SU(2)_{k_2}$
gauged supergravity \cite{CSF}. This can be easily shown in the 2-d
$\sigma$-model approach by means of the $\alpha'$-expansion. More
precisely, the gauging is:
$$ [SU(2)_{k_1}\otimes SU(2)_{k_2}]_{\rm left} \otimes
[SU(2)_{k_1}\otimes SU(2)_{k_2}]_{\rm right} $$
in type-II construction ($N=8$ gauged supergravity), and
$$ [SU(2)_{k_1}\otimes SU(2)_{k_2}]_{\rm left} \otimes
[SU(2)_{k_1}\otimes SU(2)_{k_2}]_{\rm right}\otimes G $$
in heterotic construction ($N=4$ gauged supergravity). The gauge
group $G$ depends on the particular embedding in the 10-d $SO(32)$ or
$E_8\otimes E_8$ gauge group, leading for instance to $G= E_7\otimes
E_8$ level-1, etc.

This connection is very important because it allows the derivation of
the 4-d
effective supergravity action, up to two space-time derivatives,
which is induced by the ${\bf K}^{(6)}$ compactification. In the
context of 2-d $\sigma$-model, it implies the knowledge of the
corresponding $\beta$-functions of the background fields. For
instance in the heterotic case, the induced $N=4$ supergravity is
uniquely fixed in terms of the heterotic gauge group mentioned
above \cite{Roo}.
The strength of the gauge couplings is also fixed by the levels of
the affine algebras, $g_A=1/\sqrt{k_A}$. The bosonic part of the
$N=4$
supergravity action, restricted in the supergravity multiplet,
reads \cite{CSF},\cite{Roo} :
\begin{equation}
S^{\rm eff}_{\rm bos}=\int d^4 x{\sqrt {-g}} \big\{ \frac{1}{2}{\cal
R}-(\nabla\Phi)^2- e^{4\Phi}\ (\nabla a)^2 -
\frac{1}{2g_A^2}(e^{-2\Phi}{\rm Tr} F^{A}\cdot F^{A} + a\ {\rm Tr}
F^{A}\cdot{\tilde F}^{A}) -V(\Phi ) \big\}
\label{Seff}
\end{equation}
where $\Phi$ is the dilaton field, $a$ is the pseudoscalar axion
(dual to the two-index antisymmetric tensor), $F^{A}_{\mu\nu}$ are
the $SU(2)_{k_A}$ field strengths, and ${\tilde F}^{A}_{\mu\nu}$ are
their duals. The potential $V(\Phi )$ is non-vanishing, as expected
from the $\sigma$-model evaluation of the dilaton $\beta$-function
\cite{betafunct}.
In fact, the ${\bf K}^{(6)}$ compactification gives rise to a
non-zero curvature contribution, inducing a non-trivial dilaton
potential proportional to the central charge deficit, $\delta{\hat
c}$ \cite{ABS}:
\begin{eqnarray}
V(\Phi ) &=&\frac{1}{2}\delta{\hat c}\ e^{2\phi}\nonumber \\
\delta{\hat c} &=& \frac{2}{3}\left(\frac{3k_1}{k_1+2}
+\frac{3k_2}{k_2+2}
-6 \right)\nonumber \\
&=& -4 \left(\frac{1}{k_1+2} +\frac{1}{k_2+2}\right)\simeq
-4(g_1^2+g_2^2) +
{\cal O}\left(\frac{1}{k_A^2}\right)\ .
\label{Vertphi}
\end{eqnarray}
It follows that the large $k_A$ limit of (\ref{Vertphi}) reproduces
the potential of the $N=4$ gauged supergravity. The ${\cal
O}(1/k_A^2)$ corrections are due to curvature effects which are
related to higher derivative terms neglected in the effective action.

\subsection{Non-critical superstrings}

The type-A, -B and -C constructions based on $W_{k_A}^{(4)}$,
and
$C_{k_A}^{(4)}$ superconformal systems
are strongly connected to the non-critical superstrings in the
so-called strong coupling regime ($1\leq \hat {c}_{matter} \leq 9$)
\cite{noncrit1}$-$\cite{2dgrav}. In
fact, the Liouville superfield of non-critical strings can be
identified with the supercoordinate of the above spaces, which has a
non-zero background charge. The central charge of the Liouville
supercoordinate can be easily determined:
\begin{equation}
{\hat c}_L=1+2(Q_1^2+Q_2^2)= 1 + 4\left(\frac{1}{k_1+2}
+\frac{1}{k_2+2}\right)
\ ,
\label{cliou}
\end{equation}
where we have used the relation among the levels $k_A$ and the
background charges $Q_A$, $Q_A^2=2/(k_A+2)$. As we will see in the
next Section, this relation follows from the $N=4$ superconformal
symmetry in both $W$ and $C$ systems. The remaining matter part
consists of tensor products of unitary $N=1$ superconformal systems
based on $SU(2)_{k_A}$ WZW, $[SU(2)/U(1)]_{k_A}$ GKO cosets, as well
as $U(1)$ factors. The matter central charge is always
\begin{equation}
{\hat c}_M=9 - 4\left(\frac{1}{k_1+2} +\frac{1}{k_2+2}\right)\ ,
\label{cmat}
\end{equation}
and it varies in the region $5\leq{\hat c}_M\leq 9$. Thus, our
explicit constructions show the existence of super-Liouville theories
coupled to $N=1$ superconformal unitary matter systems in the strong
coupling regime. The problematic complex conformal weights, usually
present in this regime, are projected out by the ($N=4$)-induced
generalized GSO projection (see Section 4). This projection
phenomenon is similar to the one observed in ref.\cite{gervais}
for the ${\hat c}_M=5$ case and in ref.\cite{{kuseiberg}}
for the
case of the $N=2$ globally defined superconformal symmetry.

The lower value ${\hat c}_M=5$ corresponds to the type-A$(i)$
construction
of (\ref{constr}), in the limiting case where $k_1=k_2=0$. In this
limit, the bosonic $SU(2)_{k_1}\otimes SU(2)_{k_2}$ currents decouple
and only their free-fermionic superpartners remain which form a
${\hat c}=2$, $N=1$ superconformal system. This subsystem, combined
with the two free supercoordinates $F^{(2)}$ in (\ref{constr}) and
with the linear combination of the two $U(1)$'s with no background
charge, define all together the ${\hat c}_M=5$ unitary matter system.
It was argued that this value of ${\hat c}_M$ is special in the sense
that it is the only super-Liouville theory with massless excitation
\cite{kuseiberg},
and in that respect it has a behaviour similar to that in
the $c_M=1$ bosonic
case. Furthermore, ${\hat c}_M=5$ is one of the special
dimensions of super-Liouville theories studied by Bilal and Gervais in
ref.\cite{gervais}. The ${\hat c}_M=5$ system
will be further investigated in Section 4.

Another interesting value is ${\hat c}_M=7$, obtained in particular
when $k_1=0$ and $k_2\rightarrow\infty$, or when $k_1=k_2=2$. It
turns out that this case corresponds to the high-temperature phase of
the heterotic critical string \cite{highT}.

\section{Exact realizations of $N=4$, ${\hat c}=4$
superconformal algebra}
\setcounter{equation}{0}
The desired $N=4$ superconformal algebra involves operators of
conformal
weights 2, 3/2, and 1, namely the stress-energy tensor $T(z)$, four
supercurrents $G_a(z)$, $a=1,2,3,4$, and three $SU(2)_n$ level-$n$
Kac-Moody currents $S_i(z)$, $i=1,2,3$. The closure of the algebra
implies the following OPE relations among these operators:
\begin{eqnarray}
T(z)T(w) &{\sim}& \frac{3\hat{c}}{4(z-w)^{4}}+\frac{2T(w)}{(z-w)^2}+
\frac{\partial T(w)}{(z-w)}\nonumber \\
T(z)G_a(w) &{\sim}& \frac{3G_a(w)}{2(z-w)^2}+
\frac{\partial G_a(w)}{(z-w)}\nonumber \\
T(z)S_i(w) &{\sim}& \frac{S_i(w)}{(z-w)^2}+
\frac{\partial S_i(w)}{(z-w)}\nonumber \\
G_{4}(z)G_{4}(w) &{\sim}& \frac{\hat{c}}{(z-w)^3}
+\frac{2T(w)}{(z-w)}\nonumber \\
G_{i}(z)G_{j}(w) &{\sim}& \delta_{ij}\frac{\hat{c}}{(z-w)^3}
-4\epsilon_{ijl}\frac{S_{l}(w)}{(z-w)^2}+
2\delta_{ij}\frac{T(w)}{(z-w)}\nonumber \\
G_{4}(z)G_{i}(w) &{\sim}& \frac{4S_{i}(w)}{(z-w)}\nonumber \\
S_{i}(z)G_{4}(w) &{\sim}& -\frac{G_{i}(w)}{2(z-w)}\nonumber \\
S_{i}(z)G_{j}(w) &{\sim}& \frac{1}{2(z-w)}\left(\delta_{ij}G_{4}(w)
+\epsilon_{ijl}G_{l}(w)\right)\nonumber \\
S_{i}(z)S_{j}(w) &{\sim}& -\delta_{ij}\frac{n}{2(z-w)^2}
+\epsilon_{ijl}\frac{S_{l}(w)}{(z-w)}\ .
\label{ope}
\end{eqnarray}
The central charge $\hat{c}$ and the level $n$ of the $SU(2)_n$
currents are related by $\hat{c}=4n$. The condition $\hat{c}=4$
implies the existence of $SU(2)_1$ currents.

Below, we present the four different realizations of the algebra,
used to define the $N=4$ superconformal blocks that appear in
(\ref{constr}).

\subsection{The free-field realization}

In this realization \cite{ademolo}, the $N=4$ basic operators are
constructed in terms of four free supercoordinates ($\Phi_a,\Psi_a$).
The $U(1)^4$ bosonic currents $J_a=\partial\Phi_a$ and the
Weyl-Majorana
2-d fermions $\Psi_a$ are normalized as:
\begin{eqnarray}
J_{a}(z)J_{b}(w) &{\sim}& -\frac{\delta_{ab}}{(z-w)^{2}}\nonumber \\
\Psi_ {a}(z)\Psi_{b}(w) &{\sim}& -\frac{\delta_{ab}}{(z-w)}\ .
\label{freeope}
\end{eqnarray}
The generators of the algebra are then given as:
\begin{eqnarray}
T &=& -\frac{1}{2}\left(J^{2}_{a}-\Psi_{a}\partial\Psi_{a}\right)
\nonumber \\
G_{1} &=& +J_{1}\Psi_{1}+J_{2}\Psi_{2}+J_{3}\Psi_{3}+J_{4}\Psi_{4}
\nonumber \\
G_{2} &=& +J_{1}\Psi_{2}-J_{2}\Psi_{1}-J_{3}\Psi_{4}+J_{4}\Psi_{3}
\nonumber \\
G_{3} &=& -J_{1}\Psi_{4}+J_{2}\Psi_{3}- J_{3}\Psi_{2}+J_{4}\Psi_{1}
\nonumber \\
G_{4} &=& -J_{1}\Psi_{3}-J_{2}\Psi_{4}+J_{3}\Psi_{1}+J_{4}\Psi_{2}
\nonumber \\
S_{i} &=& \frac{1}{2}\left(\Psi_{4}\Psi_{i}+
\frac{1}{2}\epsilon_{ijl}\Psi_{j}\Psi_{l}\right)\ .
\label{foper}
\end{eqnarray}

For later convenience, it is useful to complexify the coordinate
currents as:
\begin{eqnarray}
P &=& J_{1}+iJ_{2} \, , \;\;\;\;\;  P^{\dag}=-J_{1}+iJ_{2}~,
\nonumber \\
\Pi &=& J_{4}+iJ_{3} \, , \;\;\;\;\;  \Pi^{\dag}=-J_{4}+iJ_{3}~,
\label{complb}
\end{eqnarray}
with
\begin{eqnarray}
P(z)P^{\dag}(w) &{\sim}& \frac{2}{(z-w)^{2}}+2T_{P}
\nonumber \\
\Pi(z)\Pi^{\dag}(w) &{\sim}& \frac{2}{(z-w)^{2}}+2T_{\Pi}\ ,
\label{compldag}
\end{eqnarray}
where $T_{P}$ and $T_{\Pi}$ are the stress tensor of the
($P,P^{\dag}$)
and ($\Pi,\Pi^{\dag}$) conformal sub-systems.

It is also  useful to bosonize the free fermions
in terms of two bosons, $H^{+}$ and $H^{-}$. First, we decompose the
$SO(4)_{1}$ level-1 fermionic currents, $\Psi_{i}\Psi_{j}$, in
terms of two $SU(2)_1$ currents, $S_{i}$, $\tilde{S}_{i}$ using the
$SO(4)$ self-dual and anti-self-dual decomposition:
\begin{eqnarray}
S_{i} &=& \frac{1}{2}\left(+\Psi_{4}\Psi_{i}+
\frac{1}{2}\epsilon_{ijl}\Psi_{j}\Psi_{l}\right)\ \
\rightarrow\ \ \left( \frac{1}{2} \partial H^{+} \, , \;\;
e^{ \pm i \sqrt{2} H^{+}}\right)\nonumber \\
\tilde{S}_{i} &=& \frac{1}{2}\left(-\Psi_{4}\Psi_{i}+
\frac{1}{2}\epsilon_{ijl}\Psi_{j} \Psi_{l}\right)\ \
\rightarrow\ \ \left( \frac{1}{2} \partial H^{-} \, , \;\;
e^{ \pm i \sqrt{2} H^{-}}\right)\, .
\label{selfdu}
\end{eqnarray}
In the above equation, $S_{i}$ and $\tilde{S}_{i}$ are parametrized
in terms of the two free bosons, $H^{+}$ and $H^{-}$, which are both
compactified on a cycle with radius $R_{H^{+}}=R_{H^{-}}= \sqrt{2}$
(the self-dual extended symmetry points).

In terms of $P,P^{\dag}$, $\Pi ,\Pi^{\dag}$,
$H^{+}$ and $H^{-}$, the $N=4$ operators become:
\begin{eqnarray}
T &=& - \frac{1}{2} \left((\partial H^{+})^{2}+
(\partial H^{-})^{2}-PP^{\dag} -\Pi \Pi^{\dag} \right)\nonumber \\
{G} &=& -\left( \Pi^{\dag}e^{-\frac{i}{\sqrt{2}}H^{-}} + P^{\dag}e^{+
\frac{i}{\sqrt{2}} H^{-}} \right) e^{+ \frac{i}{ \sqrt{2}}H^{+}}
\nonumber \\
{\tilde {G}} &=& \left( \Pi \;\; e^{+\frac{i}{\sqrt{2}}H^{-}} - P\;\;
e^{-\frac{i}{ \sqrt{2}} H^{-}} \right) e^{+ \frac{i}{\sqrt{2}}H^{+}}
\nonumber \\
S_{3} &=& \frac{1}{\sqrt{2}} \partial H^+~,
{}~~S_{\pm} = e^{\pm i\sqrt{2}H^+}\ ,
\label{complexsup}
\end{eqnarray}
where
$$
G = \frac{G_{1}+iG_{2}}{\sqrt{2}} \, , \;\;\;\;\;\;\;\;
{\tilde {G}}= \frac{G_{4}+iG_{3}}{\sqrt{2}}\, .
$$

The above expressions clearly show  that the supercurrents
(${G},G^{\dag}$) and (${\tilde {G}},{\tilde {G}}^{\dag}$) form two
doublets under $SU(2)_{H^{+}}$. The $H^{+}$ factorization in the
supercurrents is not a particular property of the free-field
realization, but a generic property for any ${\hat{c}}=4$ system
with $N=4$ symmetry. Consequently, the supercurrents always have
a factorized  product form in terms of two conformal
operators. The first one is not dependent on $H^{+}$ and has
conformal
dimension $\frac{5}{4}$, while the other is given only in terms of
$H^{+}$ and has conformal dimension $\frac{1}{4}$. On the other hand,
${G}$ and ${\tilde {G}}$ do not transform covariantly under the
action of $SU(2)_{H^{-}}$. They are odd, however, under a ${\bf Z}_2$
transformation, defined by $(-)^{2{\tilde S}}$, which is the parity
operator associated to the $SU(2)_{H^{-}}$ spin ${\tilde S}$ (integer
spin representations are even, while half-integer representations are
odd).

Another useful global quantum number is obtained by combining the
bosonic oscillator number $N_{P}$, which counts the number of
the $P$-oscillators minus the number of  $P^{\dag}$-ones, with the
$SU(2)_{H^{-}}$
helicity ${\tilde S}_3$:
\begin{equation}
{\cal N}^-\equiv N_{P}+{\tilde S}_3\ .
\label{NPnumber}
\end{equation}
${G}$ and ${\tilde {G}}$ have $\mp 1/ 2~~$  (${\cal N}^-$)-charges,
respectively. As we will see in Section 4,
the ${\cal N}^-$ charge, the $(-)^{2{\tilde S}}$ parity, as well as
the $SU(2)_{H^{+}}$ spin $S$ play an important role in the
definition of the induced $N=4$ generalized GSO projections.

\subsection{The semi-wormhole
realization}

In this case, the basic operators of the $N=4$ algebra are
constructed in terms of the $SU(2)_k\otimes U(1)_Q$ bosonic currents
$J_a$ and their free-fermionic superpartners $\Psi^a$
\cite{kprn4},\cite{n4kounnas}:
\begin{eqnarray}
T &=& -\frac{1}{2}\left[ \frac{2}{k+2}J_i^2 + J_4^2
-\Psi_a\partial\Psi_a + Q\partial J_4\right]\nonumber \\
G_4 &=& \sqrt{\frac{2}{k+2}}\left( J_i\Psi_i+
\frac{1}{3}\epsilon_{ijl}\Psi_i\Psi_j\Psi_l\right) + J_4\Psi_4
+Q\partial\Psi_4\nonumber \\
G_i &=& \sqrt{\frac{2}{k+2}}\left( -J_i\Psi_4+\epsilon_{ijl}J_j\Psi_l
-\epsilon_{ijl}\Psi_4\Psi_j\Psi_l\right) + J_4\Psi_i
+Q\partial\Psi_i\nonumber \\
S_{i} &=& \frac{1}{2}\left(\Psi_{4}\Psi_{i}+
\frac{1}{2}\epsilon_{ijl}\Psi_{j}\Psi_{l}\right)\ .
\label{semiwdef}
\end{eqnarray}
The closure of the $N=4$ algebra can be easily verified using the OPE
relations among $J_a$ and $\Psi_a$:
\begin{eqnarray}
J_4(z)J_4(w) &{\sim}& -\frac{1}{(z-w)^{2}}\nonumber \\
J_{i}(z)J_{j}(w) &{\sim}& -\frac{k}{2}\frac{\delta_{ij}}{(z-w)^{2}}
+\epsilon_{ijl}\frac{J_l}{(z-w)}\nonumber \\
\Psi_{a}(z)\Psi_{b}(w) &{\sim}& -\frac{\delta_{ab}}{(z-w)}\ .
\label{Wope}
\end{eqnarray}
The relation $Q=\sqrt{\frac{2}{k+2}}$ between the background charge
$Q$ and the level $k$ is necessary for the cancellation of the
central charge deficit $\delta{\hat c}=-4/(k+2)$, induced by the
non-flat $S^3$ subspace, by the central charge benefit $2Q^2$,
induced by the non-trivial dilaton. Indeed, the presence of a
non-zero background charge for the $J_4$ coordinate current implies,
in the $\sigma$-model representation, a dilaton background linear in
the fourth coordinate.

The interesting observation here is that the $S_i$ $N=4$ currents are
the same as in the free-field realization. The absence of any
curvature correction ${\cal O}(1/k)$ is due to an exact cancellation
among the contribution of the torsion terms $\Psi_a\Psi_b\Psi_c$ and
the contribution of the background charge terms $Q\partial\Psi_a$
appearing in the $N=4$ supercurrents. This cancellation is a
consequence of the $N=4$ algebra with ${\hat c}=4$, which implies the
existence of an $SU(2)_1$ level-1 current algebra. In fact, in any
$N=4$ supersymmetric $\sigma$-model, the self-dual combination of the
fermionic currents formed by the fermion bilinears $\Psi_a\Psi_b$ is
always free (see (\ref{selfdu})). It happens that in the $W^{(4)}_k$
space, the anti-self-dual combination remains also free. However,
this is not true in general; the $N=4$ algebra does not forbid
non-trivial interactions of the second combination with the target
space curvature.

As in the free-field realization, it is convenient to bosonize the
free fermions by introducing the $H^+$ and $H^-$ scalar fields, as
in (\ref{selfdu}), and thus to factorize the $SU(2)_{H^+}$ dependence
of the supercurrents. After this bosonization, the $N=4$
supercurrents $G$ and ${\tilde G}$ take the same form as in
(\ref{complexsup}) in terms of the modified coordinate currents $P$
and $\Pi$:
\begin{eqnarray}
P &{\rightarrow}& P_k=Q(J_1+iJ_2)\nonumber \\
P^{\dag} &{\rightarrow}& P^{\dag}_k=Q(-J_1+iJ_2)\nonumber \\
\Pi &{\rightarrow}& \Pi_k=J_4 + iQ(J_3+\sqrt{2}\partial H^-)\nonumber
\\
\Pi^{\dag} &{\rightarrow}& \Pi^{\dag}_k=-J_4
+ iQ(J_3+\sqrt{2}\partial H^-)\ ,
\label{Pcompl}
\end{eqnarray}
while the energy-momentum tensor is shifted by the background charge:
\begin{equation}
T = -\frac{1}{2}~\bigg[ (\partial H^+)^2 + (\partial H^-)^2 + Q^2~
(J^2_1+J^2_2+J^2_3)+J^2_4+Q\partial J_4)\bigg]\ .
\label{Tcompl}
\end{equation}
The modification $Q\sqrt{2}\partial H^-$ in the expression of $\Pi$
and $\Pi^{\dag}$ gives rise at the same time to the standard
fermionic torsion terms $\pm Q\Psi_a\Psi_b\Psi_c$, and to the
fermionic background charge terms $\pm Q\partial\Psi_a$ in the
expression of the supercurrents (\ref{semiwdef}).

As in the free-field case, (${G},G^{\dag}$)
and (${\tilde {G}},{\tilde {G}}^{\dag}$) form two doublets under
$SU(2)_{H^+}$ while they are odd under the ${\bf Z}_2$ parity
$(-)^{2{\tilde S}}$. Finally, the ${\cal N}^-$ charge
(\ref{NPnumber}) is now replaced by a global $SU(2)_{k+1}$ charge
defined as the diagonal combination of $SU(2)_k$ and $SU(2)_{H^-}$:
\begin{equation}
{\cal N}_i = J_i + {\tilde S}_i\ .
\label{ninumb}
\end{equation}
Also (${G},G^{\dag}$) and (${\tilde {G}},{\tilde {G}}^{\dag}$) form
two doublets under $SU(2)_{\cal N}$. Moreover $G$ and ${\tilde {G}}$
have $({\cal N}_3, S_3)$ charges equal to $(-1/2,1/2)$ and
$(1/2,1/2)$,
respectively. As we will see in Section 4, the $SU(2)_{\cal N}$ and
$SU(2)_{H^+}$ spins play an important role as classification charges
of the string excitations around the semi-wormhole background.

The existence of non-trivial target spaces with ${\hat c}=4$ for any
value of $k$ is interesting, since it allows the study of these
models
by means of the $1/k$ expansion, where the semiclassical
approximation is valid. Indeed, the limit $k\to\infty$ can be
obtained after rescaling the $SU(2)_k$ currents
$J_i\to{\sqrt{k/2}}J_i$, and it gives back the free-field case. As we
mentioned in the introduction, the semi-wormhole interpretation of
$W^{(4)}_k$ is valid only in the large-$k$ limit. This
background interpretation fails for small $k$ (large curvature). For
instance at $k=2$ the three $SU(2)_2$ currents are equivalent to
three free world-sheet fermions, while at $k=1$ they are equivalent
to a free boson on a cycle of radius ${\sqrt 2}$.

In the $k=0$ limit, the $SU(2)_k$ currents decouple and the $N=4$
operators are described in terms of the $U(1)_Q$ current and the four
free fermions $\Psi_4$ and $\Psi_i$. From the point of view of
$N=1$ local world-sheet supersymmetry, this system is equivalent to a
free super-coordinate $(\partial X\equiv\Psi_1\Psi_2, \Psi_3)$
coupled to super-Liouville $(\partial\Phi\equiv J_4,\Psi_4)$. The
coordinate $X$ is compactified on a cycle of radius $R_X=1$, which
corresponds to the fermionization point. The heterotic superstring
solution based on $F^{(6)}\otimes W^{(4)}_{k=0}$ was identified in
ref.\cite{highT} as the high-temperature phase of the ten-dimensional
theory. There, the value of the radius of the $X$ coordinate defines
the temperature, while the special value $R_X=1$ corresponds to a
self dual thermal spectrum \cite{dualTW}, and minimizes the free
energy
\cite{highT}. In the transition from the zero-temperature phase,
described by the solution $F^{(6)}\otimes W^{(4)}_{k=\infty}$, to the
high-temperature phase ($k=0$), the decoupling of the three $SU(2)$
currents corresponds to a central charge deficit $\delta{\hat c}=2$,
which is balanced by the dilaton motion in the $k=0$ phase.

\subsection{The $C^{(4)}_k$ torus-bell realization}

In this case, the elementary fields are the
$\left( \frac{SU(2)}{U(1)}\right)_k$ parafermionic currents $P_k$ and
$P^{\dag}_k$ of conformal weight $(1-1/k)$ \cite{zamoParafemions},
two free $U(1)$ currents
$J_3=\partial X_3$ and $J_4$, and four free fermions, which are
parametrized by the $H^+$ and $H^-$ bosonic fields. The
$X_3$ coordinate is compactified on a cycle of radius $\sqrt{2k}$,
while
$J_4$ has a background charge $Q=\sqrt{2/(k+2)}$. In the large-$k$
limit, $C^{(4)}_k$ is factorized in two 2-d subspaces; the first
subspace is described by the
$SU(2)/U(1)$ bell, while the second subspace $U(1)_{X_3}\otimes
U(1)_Q$ is a two-dimensional cylinder.

The $N=4$ operators are given by \cite{n4kounnas}:
\begin{eqnarray}
T &=& - \frac{1}{2} \left((\partial H^{+})^{2}+
(\partial H^{-})^{2}+ J_3^2+ J_4^2 +Q\partial J_4 \right)
+T_{P_k}\nonumber \\
G &=& -\left( \Pi_k^{\dag}e^{-\frac{i}{\sqrt{2}}H^{-}} +
P_k^{\dag}e^{+ \frac{i}{\sqrt{2}} H^{-} -i\sqrt{\frac{2}{k}}X_3}
\right)
e^{+ \frac{i}{\sqrt{2}}H^{+}}
\nonumber \\
{\tilde {G}} &=& \left( \Pi_k \;\; e^{+\frac{i}{\sqrt{2}}H^{-}} - P_k
\;\;   e^{-\frac{i}{ \sqrt{2}} H^{-} +i\sqrt{\frac{2}{k}}X_3}
\right) e^{+ \frac{i}{\sqrt{2}}H^{+}}
\nonumber \\
S_{3} &=& \frac{1}{\sqrt{2}} \partial H^+~,
{}~~S_{\pm} = e^{\pm i\sqrt{2}H^+}\ ,
\label{cspoper}
\end{eqnarray}
where $\Pi_k$ and $\Pi_k^{\dag}$ are defined in terms of $J_3$, $J_4$
and
$H^-$:
\begin{eqnarray}
\Pi_k &=& +J_4 +i\left(\sqrt{\frac{k}{k+2}}\partial X_3
+ \sqrt{\frac{4}{k+2}}\partial H^-\right)\nonumber \\
\Pi_k^{\dag} &=& -J_4 +i\left(\sqrt{\frac{k}{k+2}}\partial X_3
+ \sqrt{\frac{4}{k+2}}\partial H^-\right)\ .
\label{Pikdef}
\end{eqnarray}
In (\ref{cspoper}), $T_{P_k}$ is the energy-momentum tensor of the
$\left( \frac{SU(2)}{U(1)}\right)_k$ parafermions with central charge
$c_P =2-6/(k+2)$. It appears in the OPE of the (non-local)
parafermionic currents $P_k$ and $P_k^{\dag}$
\begin{equation}
P_{k}(z)P_{k}^{\dag}(w) \sim
\left[\frac{k}{(k+2)}\frac{2}{(z-w)^2}+2T_{P_{k}}(w)\right]
(z-w)^{\frac{2}{k}}\ .
\label{Tparaf}
\end{equation}
In the supercurrent expression (\ref{cspoper}), the deviation from
the free-field dimensionality $-1/k$ of the parafermionic currents
$P_k$ is cancelled by the weight $+1/k$ of the $\exp
(-i\sqrt{2/k}X_3)$ modification.

In contrast to the semi-wormhole case, there is no $SU(2)_{\cal N}$
globally defined charge (\ref{ninumb}). Instead, there is only an
abelian charge ${\cal N}_3$ defined by:
\begin{equation}
{\cal N}_3 = \sqrt{\frac{k}{2}}J_3 + {\tilde S}_3\ ;
\label{nthreen}
\end{equation}
$G$ and ${\tilde G}$ have $U(1)_{{\cal N}_3}$ charges $-1/2$ and
$+1/2$,
respectively.

\subsection{The $\Delta^{(4)}_k$ cigar/trumpet-bell realization}

In this case, the elementary fields are the
$\left( \frac{SU(2)}{U(1)}\right)_{k}$ (compact) parafermionic
currents \cite{zamoParafemions} $P_k$ and $P_k^{\dag}$, the $\left(
\frac{SL(2,R)}{U(1)}\right)_{k'}$ non-compact parafermionic currents
\cite{DLPnoncompact}
${\hat\Pi}_{k'}$ and ${\hat\Pi}_{k'}^{\dag}$ of conformal weight
$(1+1/k')$, as well as the two free bosons $H^+$ and $H^-$. The level
$k'=k+4$, so that the total central charge ${\hat c}$ remains equal
to 4 for any value of $k$. In contrast with the three previous cases,
the compactification radius of $H^-$ deviates from its self-dual
value due to $1/k$ background curvature corrections.
Its exact value is $R_{H^-}=\sqrt{2}\sqrt{k'/k}$ and, thus, breaks
$SU(2)_{H^-}$ to a $U(1)$ subgroup. However, as we have already
mentioned, $H^+$ remains intact as a consequence of the
$N=4$ algebra with ${\hat c}=4$. For large $k$, $\Delta^{(4)}_k$ is
factorized in two 2-d subspaces; the fisrt one is the $SU(2)/U(1)$
bell, while the second one is described by the $SL(2,R)/U(1)$ cigar
(axial gauging) or trumpet (vector gauging) \cite{giveondual}.

The $N=4$ operators are given by \cite{n4kounnas}:
\begin{eqnarray}
T &=& - \frac{1}{2} \left((\partial H^{+})^{2}+(\partial H^{-})^{2}
 \right)+T_{P_{k}}+T_{{\hat\Pi}_{k'}}
\nonumber \\
G &=& -\left( {\hat\Pi}^{\dag}_{k'} e^{-i\sqrt{\frac{k}{2(k+4)}}
H^{-}} + P^{\dag}_{k} e^{+ i\sqrt{\frac{k+4}{2k}} H^{-}}
\right) e^{+ \frac{i}{\sqrt{2}}H^{+}}
\nonumber \\
{\tilde {G}} &=& \left( {\hat\Pi}_{k'} \;\;
e^{+i\sqrt{\frac{k}{2(k+4)}}
H^{-}} - P_{k} \;\; e^{- i\sqrt{\frac{k+4}{2k}} H^{-}}
\right) e^{+ \frac{i}{\sqrt{2}}H^{+}}
\nonumber \\
S_{3} &=& \frac{1}{\sqrt{2}} \partial H^+~,
{}~~S_{\pm} = e^{\pm i\sqrt{2}H^+}\ ,
\label{dspoper}
\end{eqnarray}
where $T_{{\hat\Pi}_{k'}}$ is the energy-momentum tensor of the
$\left( \frac{SL(2,R)}{U(1)}\right)_{k'}$ non-compact para- fermions
with central charge $c_{\hat\Pi}= 2+6/(k'-2)=2+6/(k+2)$. It appears
in the OPE of the non-compact parafermionic currents ${\hat\Pi}_{k'}$
and ${\hat\Pi}_{k'}^{\dag}$:
\begin{equation}
{\hat\Pi}_{k'}(z){\hat\Pi}_{k'}^{\dag}(w) \sim
\left[\frac{k'}{(k'-2)}\frac{2}{(z-w)^2}+2T_{{\hat\Pi}_{k}}(w)\right]
(z-w)^{-\frac{2}{k'}}\ .
\label{Tncparaf}
\end{equation}
In the supercurrent expression (\ref{dspoper}), the deviation from
the free-field dimensionality, $1/(k+4)$ of the non-compact
parafermionic currents ${\hat\Pi}_{k'}$ and $-1/k$ of the compact
ones $P_k$, is cancelled by the deviation from the free-fermion
dimensionality 1/2 of the exponential factors
$e^{+i\sqrt{\frac{k}{2(k+4)}}H^{-} +\frac{i}{\sqrt{2}}H^{+}}$ and
$e^{- i\sqrt{\frac{k+4}{2k}} H^{-} +\frac{i}{\sqrt{2}}H^{+}}$,
respectively. The last deviation is due to the radius modification of
$H^-$ from its self-dual value \cite{n4kounnas}.

In contrast with the three previous cases, here there is no globally
defined charge analogous to ${\cal N}$ of (\ref{NPnumber}),
(\ref{ninumb}), (\ref{nthreen}). Furthermore the ${\bf Z}_2$ parity
$(-)^{2{\tilde S}}$, under which $G$ from ${\tilde G}$ are odd, is
replaced by
\begin{equation}
e^{2i\pi kQ_{H^-}}\quad ;\quad
Q_{H^-}=\sqrt{\frac{k+4}{2k}}\oint\partial H\ ,
\label{kodd}
\end{equation}
when $k$ is odd, and $\exp~ (i\pi kQ_{H^-})$ when $k=2$ mod 4. For
$k=0$ mod 4, such a parity cannot be defined in terms of $H^-$ alone
and it is necessary to use the ${\bf Z}_k$ discrete symmetry of
compact parafermions. Under this symmetry $P_k$ picks up a phase
$\exp~(2i\pi/k)$. The generalized ${\bf Z}_k$ operator for any $k$
even is a discrete analogue of the global charge (\ref{nthreen}), and
it reads:
\begin{equation}
e^{2i\pi {\cal N}}\quad ;\quad
{\cal N}=\frac{k+2}{2}Q_{P_k} + Q_{H^-}\ ,
\label{keven}
\end{equation}
where $Q_{P_k}$ is the parafermionic charge (defined modulo integer)
and equals $2/k$ for the $P_k$ currents. One can easily check that
the $N=4$ supercurrents change sign under the above ${\bf Z}_k$
transformation.

\section{The spectrum of string excitations and partition functions}
\setcounter{equation}{0}
The explicit realizations of the $N=4$ algebra that we presented in
the previous section, in terms of known conformal field theories,
allows us to compute the spectrum of string excitations around any
of the background solutions given in (\ref{constr}). In this section,
we present the method of constructing modular-invariant combinations
respecting the $N=4$ superconformal symmetry. The latter implies
the existence of space-time supersymmetry \cite{banksdixon} in the
corresponding
non-trivial target spaces, and thus guarantees the stability of these
classical solutions in string perturbation theory. The origin of
these new target space supersymmetries follows from the world-sheet
$N=4$ spectral flow relations that imply a spectrum degeneracy among
space-time bosonic and fermionic string modes. This degeneracy
guarantees the vanishing of the vacuum energy, and thus
the background stability, at least at the one-loop level.

\subsection{The required projections}

In all constructions, the total number of space-time supersymmetries
is reduced by a factor of 2 with respect to the flat (toroidal)
compactifications. In the context of the $\sigma$-model approach, the
$N=4$ spectral flows are related to the number of covariantly
constant spinors admitted by the corresponding non-trivial target
spaces (\ref{constr}). Thus, from the point of view of four
non-compact dimensions, there exist two covariantly constant spinors
in heterotic and four in type-II backgrounds.

The reduction of space-time supersymmetries by a factor of 2 can be
easily seen in the variation of our solutions (\ref{constr}), where
the two-dimensional subspace $F^{(2)}$ is flat (non-compact), with
Lorentzian signature. The supersymmetry generators are constructed by
analytic (or antianalytic) dimension-1 currents whose transverse part
is a spin-field of dimension 1/2 constructed in terms of the $H^+_A$
and $H^-_A$ bosonized fermions (of the two ${\hat c}_A=4$ systems
$A=1,2$). In the toroidal case, there are four such spin-fields,
which
are even under the GSO parity:
\begin{equation}
e^{2i\pi (S_1 +S_2)}\ ,
\label{GSO}
\end{equation}
where $S_A$ are the two $SU(2)_{H^+_A}$ level-1 $N=4$ spins:
\begin{eqnarray}
\Theta_{\pm} &=& e^{\frac{i}{\sqrt 2}(H^+_1 \pm H^+_2)}\nonumber \\
{\rm and}~~~~\tilde{\Theta}_{\pm} &=& e^{\frac{i}{\sqrt 2}(H^-_1 \pm
H^-_2)}\ .
\label{susy}
\end{eqnarray}
In the case of the non-trivial spaces described in Section 3, only
the two supersymmetry generators based on the operators
$\Theta_{\pm}$, which are constructed with $H^+$'s, are
BRS-invariant. Indeed, the other two operators $\tilde{\Theta}_{\pm}$
do
not exist in the free-field (${\bf Z}_2$ orbifold) and the $C^{(4)}$
realization; in the former case, this is because they are not ${\bf
Z}_2$-invariant, while in the latter case, because $H^-$ is not
compactified at the self-dual point. Moreover in
$W^{(4)}$ and $C^{(4)}$ realizations, the $\tilde{\Theta}_{\pm}$
supersymmetry generators are not physical, due to the $\partial H^-$
modification, related to the torsion and/or background charge, in the
supercurrent expressions (\ref{Pcompl}) and (\ref{Pikdef}).

The global existence of the (chiral) $N=4$ superconformal algebra
implies in all our constructions a universal GSO projection that
generalizes the one of the $N=2$ algebra \cite{gepner4d},
and it is responsible for the
existence of space-time supersymmetry. This projection restricts the
physical spectrum to being odd under the total $H^+_A$ parity
(\ref{GSO}). Thus, the supersymmetry generators based on
$\Theta_{\pm}$, which are even under (\ref{GSO}), when acting on
physical states, create physical states with the same mass but with
different statistics. The GSO projection restricts
the (level-1) character combinations associated with the two
$SU(2)_{H^+}$'s to appear in the form:
\begin{equation}
\frac{1}{2} (1-(-)^{l_1+l_2})
\chi_{H^+_1}^{l_1}\chi_{H^+_2}^{l_2} =
\chi_{H^+_1}^{l_1} \chi_{H^+_2}^{1-l_1}\delta_{l_2,1-l_1}\ ,
\label{charplus}
\end{equation}
with $l_A=2S_A$ taking values 0 or 1 corresponding to the two
possible characters, (spin-$0$ and spin-$1/2$)  of the $SU(2)_{k=1}$
Kac-Moody algebras.

The above character combination is universal and is valid for all
string solutions described in Section 3. For every particular
solution, modular invariance and unitarity impose some
restrictions to the remaining  degrees of freedom. These extra
restrictions are not universal
but depend on the particular solution under consideration.

\subsection{The required characters}

The basic rules of our construction are similar to that of, the
orbifold construction \cite{orbifold}, the free 2-d
fermionic constructions \cite{abk4d},  and the Gepner construction
\cite{gepner4d} were, one combines in a modular invariant way the
world-sheet degrees of freedom consistently with unitarity and
spin-statistics of  the string spectrum. We will
choose at the beginning as first example the derivation of the string
spectrum in a $five$-$brane$ background $M^6 \otimes
W^{(4)}_k$, where $M^6=F^{(2)}\otimes F^{(4)}$ is a
six-dimensional
non-compact flat Minkowski space. The six non-compact coordinates,
together with the reparametrization ghosts ({\bf b},{\bf c}),
provide a contribution to the (type-II) partition function:
\begin{equation}
Z_{B}[F^{(6)};({\bf b},{\bf c})]= \frac{{\rm Im}\tau^{-2}}{\eta ^{4}
(\tau)\bar{\eta}^{4}(\bar{\tau})}\ .
\label{zbf6}
\end{equation}

The contribution of the $M^6$  world-sheet fermions together with
the ${\bf \beta}$ and ${\bf \gamma}$ super-reparametrization ghosts
is:
\begin{equation}
Z_{F}[M^6;({\bf \beta},{\bf \gamma})]= (-)^{\alpha +\beta}
\frac{\theta ^{2} (^{\alpha}_{\beta})}{ \eta ^{2}(\tau)}
(-)^{\bar{\alpha} +\bar{\beta}}
\frac{\bar{\theta}^{2}(^{\bar{\alpha}}_{\bar{\beta}})}
{\bar{\eta}^{2}(\bar{\tau})} \ ,
\label{zff6}
\end{equation}
where $\alpha$ and $\beta$ denote the spin structures. In
(\ref{zff6}), the spin-statistic factor $(-)^{\alpha +\beta}$ and
$(-)^{\bar{\alpha} +\bar{\beta}}$ comes from the contribution of the
(left- and right-moving) $F^{(2)}$ world-sheet fermions and the
(left- and right-moving) $({\bf\beta}, {\bf\gamma})$-ghosts.
The Neveu-Schwarz ($NS$, $\overline{NS}$) sectors correspond to
$\alpha, \bar{\alpha}=0$ and the Ramond ($R,\bar{R}$) sectors
correspond to $\alpha, \bar{\alpha}=1$. For later convenience we
decompose the $O(4)$ level-1 characters, which are written with
theta-functions, in terms of the $SU(2)_{H^{+}_{1}}\otimes
SU(2)_{H^{-}_{1}}$ characters using the identity:
\begin{equation}
\frac{\theta ^{2} (^{\alpha}_{\beta})}{ \eta ^{2}(\tau)}=
\sum_{l=0}^{1} (-)^{\beta l} \chi^{l}_{H^{+}_1}
\chi^{l+\alpha(1-2l)}_{H^{-}_1}\ ,
\label{theeta}
\end{equation}
and similarly for the right-movers. The modular
transformations of $\theta$-functions are:
\begin{eqnarray}
\tau\rightarrow\tau +1~~~&:&~~~
\frac{\theta^{2}(^{\alpha}_{\beta})}{\eta^2} \longrightarrow
e^{i\pi\frac{\alpha^2}{2} -{i\pi\over 6}}
\frac{\theta ^{2}(^{~~\alpha}_{\beta+\alpha+1})}{\eta^2}
\nonumber \\
\tau\rightarrow{-1/\tau}~~~&:&~~~
\frac{\theta^{2}(^{\alpha}_{\beta})}{\eta^2} \longrightarrow
e^{i\pi\alpha\beta} \frac{\theta^{2}(^{\beta}_{\alpha})}{\eta^2}\ .
\label{thetatransf}
\end{eqnarray}

Then, one must combine the above $M^6$ characters with those of
$W^{(4)}_k$, namely: (i) the $SU(2)_k$ ($\chi ^{L}_k~,~L=1,2,\cdots,
k$), (ii) the $U(1)_Q$ Liouville-like ones, (iii) the
$SU(2)_{H^{+}_{2}}$ ($\chi ^{l}_{H^{+}_{2}}~,~ l=0,1 $), and (iv)
the $SU(2)_{H^{-}_{2}}$ ($\chi ^{l}_{H^{-}_{2}}~,~ l=0,1 $).

The $U(1)_Q$ Liouville-like characters can be classified in two
categories. Those that correspond to the continuous representations
generated by the lowest-weight operators:
\begin{equation}
e^{\beta X_L}\ ;\quad \beta=-\frac{1}{2}Q +ip\ ,
\label{contchar}
\end{equation}
having positive conformal weights $h_p=\frac{Q^2}{8}+\frac{p^2}{2}$.
The fixed imaginary part in the momentum $iQ/2$ of the plane waves,
is due to the non-trivial dilaton motion. The second category of
Liouville characters corresponds to lowest-weight operators
(\ref{contchar}) with $\beta=Q\tilde{\beta}$ real, leading to
negative conformal dimensions
$-\frac{1}{2}\tilde{\beta}(\tilde{\beta}+1)Q^2=
-\frac{\tilde{\beta}(\tilde{\beta}+1)}{k+2}$. Both categories of
Liouville representations give rise to unitary representations of
the $N=4$, ${\hat c}=4$ system $W^{(4)}_k$, once they are combined
with
the remaining degrees of freedom. The continuous representations
(\ref{contchar}) form long (massive) representations \cite{ademolo}
of $N=4$ with
conformal weights larger than the $SU(2)$ spin, $h>S$. On the
other hand, the second category contains short representations of
$N=4$ \cite{ademolo}
($h=S$), while $\beta$ can take only a finite number of values,
$-(k+2)/2\le\tilde{\beta}\le k/2$. In fact, their locality with
respect to the
$N=4$  operators implies:
\begin{eqnarray}
S &=& \frac{1}{2},\quad \tilde{S}=\frac{1}{2}:\quad \tilde{\beta} =
-(j+1) \nonumber \\
S &=& 0,\quad \tilde{S}=0:\quad \tilde{\beta} = j\ .
\label{discrchar}
\end{eqnarray}

In both cases of (\ref{discrchar}), the conformal weight $h=S$ is
independent of $SU(2)_k$ and $SU(2)_{H^-}$ spins, due to the
cancellation between the Liouville and $SU(2)_k$ contributions. The
states associated to the short representations of $N=4$ do not have
the interpretation of plane waves, but they describe a discrete set
of
bound states. They are similar to the discrete states found in the
$c=1$ matter system coupled to the Liouville field and also to the
two-dimensional coset models \cite{blackholever}. Although they
play a crucial role
in scattering amplitudes, they do not correspond to asymptotic
states and nor do they  contribute to the partition function. Indeed
in our case they are not only discrete but also
finite in number and thus, have zero measure compared to the
contribution of continuous representations.

The presence of discrete
representations with $\beta$ positive are necessary to define
correlation functions. In fact, the balance of the background
charge for an $N$-point amplitude at genus $g$ implies the relation
\begin{equation}
N+2(g-1)+2\sum_I\tilde{\beta}_I=0\ ,
\label{screen}
\end{equation}
where the sum is extended over the vertices of the discrete
representation states. Thus, these vertices define an appropriate set
of screening operators, necessary to define amplitudes in the
presence
of non-vanishing background charge. In our case the screening
procedure has an interesting physical interpretation similar to the
scattering of asymptotic propagating states (continuous
representations) in the presence of non-propagating bound states
(discrete representations). The screening operation then describes
the
possible angular momentum excitations of the bound states. Below, we
restrict ourselves to the one-loop partition function, where the
discrete representations are not necessary (see eq.(\ref{screen})
with
$g=1$ and $N=0$).

It is convenient to define appropriate character combinations of
$SU(2)_k$, which transform covariantly under modular transformations:
\begin{equation}
Z_k[^{\alpha}_{\beta}] = \sum_{L=0}^k e^{i\pi\beta L} \chi_k^L
\bar{\chi}_k^{L+\alpha (k-2L)} \ ,
\label{zab}
\end{equation}
where $\alpha$, $\beta$ can be either 0 or 1. Under modular
transformations, $Z_k[^{\alpha}_{\beta}]$ transforms as:
\begin{eqnarray}
\tau\rightarrow\tau +1~~~:~~~
&Z_k[^{\alpha}_{\beta}]& \longrightarrow
e^{-i\pi\frac{k}{2}\alpha^2} Z_k[^{~\alpha}_{\beta+\alpha}]
\nonumber \\
\tau\rightarrow{-1/\tau}~~~:~~~
&Z_k[^{\alpha}_{\beta}]& \longrightarrow
e^{i\pi k\alpha\beta} Z_k[^{\beta}_{\alpha}]\ .
\label{zabtransf}
\end{eqnarray}
Because of the $k$-dependent phase in the $\tau\rightarrow\tau +1$
transformation, we must distinguish four different cases
corresponding to
$k=0,1,2,3$ modulo 4.

The partition function must satisfy two basic constraints emerging
from the $N=4$ algebra. The first is associated to the two spectral
flows of the $N=4$ algebra which impose the universal GSO projection
(\ref{GSO}), (\ref{charplus}) of the $H^+$ spin. The second
constraint is associated to the reduction of space-time
supersymmetries by a factor of 2. It imposes a second projection
which eliminates half of the lower-lying states constructed with the
$H^-_2$ field. Indeed, the vertex operators
\begin{equation}
e^{ip_{\mu}X^{\mu}+(ip -{Q\over 2})X_L}
e^{\frac{i}{\sqrt 2}(H^+_{1,2} \pm H^-_1)}
= e^{ip_{\mu}X^{\mu}+(ip -{Q\over 2})X_L}\
[({\rm spin}{\Psi}^{\mu})_+~{\rm or}~({\rm spin}{\Psi}^{I})_+]\
({\rm spin} {\Psi}^{\mu})_-
\label{statesp}
\end{equation}
create physical states (bosons and fermions correspond to $H^+_1$ and
$H^+_2$, respectively), while the vertex operators
\begin{equation}
e^{ip_{\mu}X^{\mu}+(ip -{Q\over 2})X_L}
e^{\frac{i}{\sqrt 2}(H^+_{1,2} \pm H^-_2)}=
e^{ip_{\mu}X^{\mu}+(ip -{Q\over 2})X_L}\
[({\rm spin}{\Psi}^{\mu})_+~{\rm or}~({\rm spin}{\Psi}^{I})_+]\
({\rm spin}{\Psi}^{I})_-
\label{statesunp}
\end{equation}
are unphysical, since they  are not local with respect to the
$N=4$ generators. These unphysical states should be eliminated from
the spectrum by additional GSO projection(s). In (\ref{statesp}) and
(\ref{statesunp}), $({\rm spin}\Psi^{\mu})_{\pm}$ and $({\rm
spin}\Psi^{I})_{\pm}$ are the spin-fields of $SU(2)_{H^{\pm}_1}$ and
$SU(2)_{H^{\pm}_2}$, respectively.

\subsection{Partition function for the wormhole solution}

For $k$ even, there is a $Z_2$ automorphism of $SU(2)_k$ which leaves
invariant the currents but acts non-trivially on the spinorial and
odd
spin representations. This allows to correlate the $SU(2)_{H_2^-}$
and
$SU(2)_k$ spins in a way which projects out of the spectrum the
unphysical states (\ref{statesunp}). A modular-invariant partition
function with this property is:
\begin{eqnarray}
Z_W = {{\rm Im}~\tau^{-5/2}\over \eta^5{\bar\eta}^5} & &{1\over 8}
\sum_{\alpha,\beta,{\bar\alpha},{\bar\beta},\gamma,\delta}
(-)^{\alpha+\beta}
\frac{\theta^{2}(^{\alpha}_{\beta})}{\eta^2}
\frac{\theta^{2}(^{\alpha+\gamma}_{\beta+\delta})}{\eta^2}
\nonumber \\
& & (-)^{{\bar\alpha}+{\bar\beta}}
\frac{{\bar\theta}^{2}(^{\bar\alpha}_{\bar\beta})}{{\bar\eta}^2}
\frac{{\bar\theta}^{2}
(^{{\bar\alpha}+\gamma}_{{\bar\beta}+\delta})}{{\bar\eta}^2}
(-)^{\delta(\alpha+{\bar\alpha}+{k\over 2}\gamma)}
Z_k[^{\gamma}_{\delta}]\ ,
\label{pf}
\end{eqnarray}
where the first factor in the r.h.s. represents the contribution of
the non-compact coordinates (\ref{zbf6}) together with the
contribution of the Liouville mode. Using the expressions
(\ref{theeta}) and (\ref{zab}), one can decompose the above partition
function in terms of $SU(2)$ characters:
\begin{eqnarray}
Z_W = {{\rm Im}~\tau^{-5/2}\over \eta^5{\bar\eta}^5}
&&\sum_{\alpha,{\bar\alpha},\gamma,l,{\bar l}}
(-)^{\alpha+{\bar\alpha}}~
\chi^{l}_{H^{+}_1} \chi^{l+\alpha}_{H^{-}_1}~
\chi^{1+l}_{H^{+}_2} \chi^{1+l+\alpha+\gamma}_{H^{-}_2}~
{\bar\chi}^{\bar l}_{H^{+}_1}
{\bar\chi}^{{\bar l}+{\bar\alpha}}_{H^{-}_1}~
{\bar\chi}^{1+{\bar l}}_{H^{+}_2}
{\bar\chi}^{1+{\bar l}+{\bar\alpha}+\gamma}_{H^{-}_2}
\nonumber \\
&&\sum_L {1\over 2}
[1+(-)^{\alpha+l+{\bar\alpha}+{\bar l}+{k\over 2}\gamma+L}]~
\chi_k^L \bar{\chi}_k^{L+\gamma (k-2L)}\ ,
\label{pfchar}
\end{eqnarray}
where the summation indices of the $SU(2)$ level-1 characters
$\chi_H$ take values modulo 2.

In the derivation of (\ref{pfchar}) from (\ref{pf}), the $\beta$ and
${\bar\beta}$ summations give rise to the universal (left- and
right-moving) GSO projections, which imply the existence of
space-time
supersymmetry. The phase $(-)^{\alpha+{\bar\alpha}}$ guarantees the
spin-statistics connection; it equals 1 for space-time bosons and
$-$1
for space-time fermions. The summation over $\delta$ gives rise to an
additional projection, which correlates the $SU(2)_{H^-_2}$ (left and
right) spin together with the spin of $SU(2)_k$ and thus reduces
the number of space-time supersymmetries by a factor of 2. This
projection takes the form:
\begin{equation}
2{\tilde S}_2 + 2{\bar{\tilde S}}_2 + L + {k\over 2}\gamma = {\rm
even}\ ,
\label{dproj}
\end{equation}
where ${\tilde S}_2$ and ${\bar{\tilde S}}_2$ are the left and right
$SU(2)_{H^-_2}$ spin. Note that $L+k/2\gamma =J+(-)^{\gamma}{\bar
J}$,
$J$ and ${\bar J}$ are the left and right $SU(2)_k$ spins.

In the $\gamma=0$ sector, the lower-lying states have (left and
right)
mass-squared $Q^2/8$ and $L=0$. Although there are seven
non-compact dimensions, there is only six-dimensional Lorentz
invariance, because of the non-trivial dilaton background. It is then
convenient to classify the states in the context of a
six-dimensional theory. In fact, the lower-lying states form the
gravitational supermultiplet of the six-dimensional $N=2$
supergravity:
\begin{equation}
(|\Psi^{\mu}> + |({\rm spin}\Psi^{\mu})_- ({\rm spin}\Psi^{I})_+>)
\otimes (|{\bar\Psi}^{\mu}> + |({\rm spin} {\bar\Psi}^{\mu})_-
({\rm spin}{\bar\Psi}^{I})_+>)~
e^{ip_{\mu}X^{\mu}+(ip -{Q\over 2})X_L}
\label{grav}
\end{equation} together with four vector multiplets:
\begin{equation}
(|\Psi^{I}> + |({\rm spin}\Psi^{\mu})_+ ({\rm spin}\Psi^{I})_->)
\otimes (|{\bar\Psi}^{I}> + |({\rm spin}{\bar\Psi}^{I})_+
({\rm spin}{\bar\Psi}^{I})_->)~
e^{ip_{\mu}X^{\mu}+(ip -{Q\over 2})X_L}
\label{vect}
\end{equation}
As expected from the effective field theory point of view,
their mass-squared $Q^2/8$ is due to the dilaton motion for bosons,
and to the non-trivial torsion for fermions.

Note that the contribution of 2-d fermions in the partition function
of the $\gamma=0$ sector is identical to the fermionic part of the
partition function of the ten-dimensional type II superstring with
an additional 1/2 factor :
\begin{equation}
Z_W^{\gamma =0} = {{\rm Im}~\tau^{-5/2}\over \eta^5{\bar\eta}^5}
{1\over 8} |\theta_3^4-\theta_4^4-\theta_2^4|^2
\sum_L |\chi_k^L|^2
\label{pfu}
\end{equation}
$Z_W^{\gamma =0}$ can be identified as the untwisted partition
function of a $Z_2$ symmetric orbifold of the ten-dimensional theory
compactified on a three-dimensional sphere $SU(2)_k$. The $Z_2$ acts
on the four (left and right) world-sheet fermions associated to the
wormhole $W^{(4)}_k$ space, as well as on the spinorial
representations of the $SU(2)_k$. This $Z_2$ projection is dictated
from the $N=4$ superconformal algebra and eliminates from the
untwisted sector the unphysical states. Modular invariance implies
the presence of a twisted sector $(\gamma=1)$, which contains states
with (left and right) mass-squared always larger than $(k-2)/16$.
The lower-lying twisted states are:
\begin{eqnarray}
(|({\rm spin}\Psi^{\mu})_+> |L=\frac{k}{2}> &+&
|({\rm spin}\Psi^{I})_+>~+~ |{\bar L}=\frac{k}{2}>) \otimes
\nonumber \\
(|({\rm spin}{\bar\Psi}^{\mu})_+> |L=\frac{k}{2}> &+&
|({\rm spin}{\bar\Psi}^{I})_+>~+~ |{\bar L}=\frac{k}{2}>)~
e^{ip_{\mu}X^{\mu}+(ip -{Q\over 2})X_L}
\label{twistat}
\end{eqnarray}
For any $k>2$, the twisted states have masses larger than $Q^2/8$
and the lower mass spectrum comes always from the $L=0$ states
contained in the untwisted sector. In that sense $k=2$ is an
exceptional case, since the lower-lying twisted states are massless
with $L=\bar{L}=k/2=1$. These states form massless unitary
representations of the $N=4$ $\hat{c}=8$ superconformal system which
saturate the $N=4$ unitarity bounds.

As we noted in Section 2.2, the above solution is connected to
non-critical superstrings with ${\hat c}_M=9-(\frac{4}{k+2})$.
Thus, for $k=2$ the matter system has ${\hat c}_M=8$ and, as we
explicitly showed above, it contains massless states in the twisted
sector. This seems to be in contradiction with the $N=2$
super-Liouville analysis of Kutasov and Seiberg \cite{kuseiberg},
where
they claim the absence of massless states for all ${\hat c}_M\ne 5$.
The reason they missed this possibility is because their analysis is
valid only for $({\hat c}_M+1)$ even integer, in order to bosonize
by pairs the $({\hat c}_M+1)$ two-dimensional fermions. In our
construction, the matter central charge is in general a fractional
number with a lower value ${\hat c}_M=23/3$ ($k=1$). In
Section 4.4 we extend our non-critical string constructions
to the general ${\bf K}^{(6)}$ space with
${\hat c}_M=9-4(\frac{1}{k_1+2}+\frac{1}{k_2+2})\ge 5$.

In the heterotic case, a modular-invariant partition function for $k$
even can be easily obtained using the heterotic map \cite{llsmap},
\cite{gepner4d}. It consists of
replacing in (\ref{pf}) the $O(4)$ characters associated to the
right-moving fermionic coordinates ${\bar\Psi}^{\mu}$, with the
characters of either $O(12)\otimes E_8$:
\begin{equation}
(-)^{{\bar\alpha}+{\bar\beta}}
\frac{{\bar\theta}^{2}(^{\bar\alpha}_{\bar\beta})}{{\bar\eta}^2}
\rightarrow
\frac{{\bar\theta}^{6}(^{\bar\alpha}_{\bar\beta})}{{\bar\eta}^6}
{1\over 2}(\frac{{\bar\theta}^{8}_3}{{\bar\eta}^8}
+ \frac{{\bar\theta}^{8}_4}{{\bar\eta}^8} +
\frac{{\bar\theta}^{8}_2}{{\bar\eta}^8} +
\frac{{\bar\theta}^{8}_1}{{\bar\eta}^8})\ ,
\label{pfheta}
\end{equation}
or $O(28)$:
\begin{equation}
(-)^{{\bar\alpha}+{\bar\beta}}
\frac{{\bar\theta}^{2}(^{\bar\alpha}_{\bar\beta})}{{\bar\eta}^2}
\rightarrow
\frac{{\bar\theta}^{14}(^{\bar\alpha}_{\bar\beta})}{{\bar\eta}^{14}}
\ .
\label{pfhetb}
\end{equation}
The lower-lying states having mass-squared $Q^2/8$ are those of a
$(4,4)$ $Z_2$ symmetric orbifold and form the spectrum of an $N=2$
six-dimensional supergravity with a gauge group either $E_7 \otimes
E_8$, or $SU(2) \otimes O(28)$.

For $k$ odd, the above constructions are not valid because there is
no $Z_2$ automorphism in $SU(2)_k$. In fact, the phase in the
$\tau\rightarrow\tau +1$ transformation (\ref{zabtransf}) cannot be
cancelled in a modular-invariant way consistently with the $N=(4,4)$
algebra, while keeping at the same time the $SU(2)_k$ left and right
symmetry. Therefore, in 6+1 non-compact dimensions, it is necessary
to twist the $SU(2)_k$ left and right currents and, thus, one must
break $SU(2)_L\otimes SU(2)_R$. However in lower non-compact
dimensions, it is possible to extend the above constructions keeping
the left and right $SU(2)_k$ symmetry. For instance, compactifying
one dimension $\phi$ on a cycle with radius $R_{\phi}={\sqrt 2}~$
(the $SU(2)$ self-dual point), one can use the additional $SU(2)$
level-one characters $\chi_{\phi}^{L'}$ to define  the
$Z_2$ action consistently.

In the case $R_{\phi}={\sqrt 2}$, a type-II modular-invariant
partition function, valid for any $k$, is:
\begin{eqnarray}
Z_W = {{\rm Im}~\tau^{-2}\over \eta^4{\bar\eta}^4}&&{1\over 8}
\sum_{\alpha,\beta,{\bar\alpha},{\bar\beta},\gamma,\delta}
(-)^{\alpha+\beta}
\frac{\theta^{2}(^{\alpha}_{\beta})}{\eta^2}
\frac{\theta^{2}(^{\alpha+\gamma}_{\beta+\delta})}{\eta^2}
\nonumber \\
&&(-)^{{\bar\alpha}+{\bar\beta}}
\frac{{\bar\theta}^{2}(^{\bar\alpha}_{\bar\beta})}{{\bar\eta}^2}
\frac{{\bar\theta}^{2}
(^{{\bar\alpha}+\gamma}_{{\bar\beta}+\delta})}{{\bar\eta}^2}
(-)^{\delta(\alpha+{\bar\alpha}+{k(k+1)\over 2}\gamma)}
Z_k[^{\gamma}_{\delta}] Z_1[^{k\gamma}_{k\delta}]\ ,
\label{pfkodd}
\end{eqnarray}
where $Z_1[^{k\gamma}_{k\delta}]$ is defined, as in (\ref{zab}), at
level-1 using the characters of the additional $SU(2)_{\phi}$
level-1 symmetry, associated to the compactified dimension $\phi$.

The partition function (\ref{pfkodd}) can be decomposed as:
\begin{eqnarray}
Z_W = &&{{\rm Im}~\tau^{-2}\over \eta^4{\bar\eta}^4}
\sum_{\alpha,{\bar\alpha},\gamma,l,{\bar l}}
(-)^{\alpha+{\bar\alpha}}~
\chi^{l}_{H^{+}_1} \chi^{l+\alpha}_{H^{-}_1}~
\chi^{1+l}_{H^{+}_2} \chi^{1+l+\alpha+\gamma}_{H^{-}_2}~
{\bar\chi}^{\bar l}_{H^{+}_1}
{\bar\chi}^{{\bar l}+{\bar\alpha}}_{H^{-}_1}~
{\bar\chi}^{1+{\bar l}}_{H^{+}_2}
{\bar\chi}^{1+{\bar l}+{\bar\alpha}+\gamma}_{H^{-}_2}
\nonumber \\
&& \sum_{L,L'} {1\over 2}
[1+(-)^{\alpha+l+{\bar\alpha}+{\bar l}+{k(k+1)\over 2}\gamma+L+kL'}]~
\chi_k^L \bar{\chi}_k^{L+\gamma (k-2L)}~
\chi_{\phi}^{L'} {\bar\chi}_{\phi}^{L'+k\gamma}\ .
\label{pfkoddchar}
\end{eqnarray}
As already mentioned above, for $k$ odd, the $Z_2$
$\delta$-projection acts also in the spinorial representations of
the additional $SU(2)_{\phi}$:
\begin{equation}
2{\tilde S}_2 + 2{\bar{\tilde S}}_2 + L + kL' + {k(k+1)\over 2}\gamma
= {\rm even}\ .
\label{dprojodd}
\end{equation}
Note that for $k$ even, this projection acts trivially on
$SU(2)_{\phi}$ representations.

As in the case of $k$-even construction, the lower lying-states in
the untwisted sector ($\gamma =0$) have mass-squared $Q^2/8$ and
$L=L'=0$. Since one of the non-compact
dimensions is compactified, we can classify these states in the
context of a five-dimensional theory; they contain the states of the
$N=4$ supergravity multiplet, together with four extra vector
multiplets.
In the twisted sector ($\gamma=1$) the states are always massive
with lower left and right mass-squared equal to
$\frac{(k+1)^2}{16(k+2)}$, corresponding to:
\vfill\eject
\begin{eqnarray}
(|({\rm spin}\Psi^{\mu},{\rm spin}\Psi^{\phi})_+ > +
|({\rm spin}\Psi^{I})_+>) &{\otimes}&
(|({\rm spin}{\bar\Psi}^{\mu},{\rm spin}{\bar\Psi}^{\phi})_+> +
|({\rm spin}\Psi^{I})_+>)
\nonumber \\
\otimes (|J={\bar J}'+{1\over 2}=\frac{k+1}{4}, J'=0,{\bar
J}'={1\over
2}> &+& |J={\bar J}'-{1\over 2}=\frac{k-1}{4}, J'=1,{\bar J}'=0>)
\nonumber \\
&& e^{ip_{\mu}X^{\mu}+(ip -{Q\over 2})X_L}\ .
\label{twistkodd}
\end{eqnarray}
For $k=1$ the twisted and untwisted lower-mass states are
degenerate; for all other values of $k$, the twisted states are
always heavier, with mass-squared larger than $Q^2/8$.

In the heterotic case, a modular-invariant partition function can be
easily obtained using the heterotic map described by (\ref{pfheta})
and (\ref{pfhetb}). The corresponding gauge groups are now
$SU(2)\otimes E_7\otimes E_8$ and $SU(2)\otimes SU(2)\otimes O(28)$,
respectively. For lower non-compact dimensions, several other
constructions can be easily obtained.

Note that in the semiclassical limit $k\rightarrow\infty$, the
twisted states become infinitely heavy, since their mass-squared
grows as $k/16$ and thus decouple from the spectrum. The remaining
states are those from the untwisted sector with lower mass-squared
$1/4k$. Moreover, the spectrum of lower modes with masses much less
than the string scale ($L<<{\sqrt k}$) forms the spectrum of the
semi-wormhole, $[L(L+2)+1]/4k$. The first term $L(L+2)/4k$ is
identical to the contribution of the angular-momentum excitations of
an ordinary Kaluza-Klein field theory  compactified on a
three-dimensional sphere, while the second term $1/4k$ is due to the
dilaton motion (or torsion). This phenomenon is similar to
the case of a toroidal compactification where the Kaluza-Klein
states have mass-squared $m^2/2R^2$ (quantized momenta) while the
stringy-like states are superheavy (in the semiclassical limit) with
mass-squared $n^2 R^2/4$. The twisted states in the wormhole space
play the same role as the winding modes in toroidal
compactifications.

The wormhole target space interpretation fails for $k$ small, as
mentioned in the introduction, since the field theory modes and
string modes have comparable masses, of the order of the string
scale. An
interesting limit is when $k=0$, which corresponds to a non-critical
superstring with ${\hat c}_M=7$. Furthermore, in the heterotic case
it corresponds to the high-temperature phase of the critical
heterotic superstring when one of the coordinates
(associated to the
temperature) is compactified on a cycle of unit radius \cite{highT}.
Unfortunately, this limit cannot be taken in
the above partition functions (\ref{pf}) and (\ref{pfkodd}), because
the $\delta$-projection is inconsistent with the global existence of
the 2-d supercurrent, when $k=0$. We were unable to find a consistent
theory with $k=0$ and more than five non-compact dimensions. When
two dimensions $\phi_{1,2}$, in addition to that of the temperature,
are also compactified on two tori of unit radius, an example of a
consistent partition function is:
\begin{eqnarray}
Z[M^4\otimes T^2\otimes W^{(4)}_{k=0}] &=&
{{\rm Im}~\tau^{-3/2}\over \eta^3{\bar\eta}^3} {1\over 8}
\sum_{\alpha,\beta,{\bar\alpha},{\bar\beta},\gamma,\delta}
(-)^{\alpha+\beta}
\frac{\theta^{2}(^{\alpha}_{\beta})}{\eta^2}
\frac{\theta^{2}(^{\alpha+\gamma}_{\beta+\delta})}{\eta^2}
\nonumber \\
& & (-)^{{\bar\alpha}+{\bar\beta}}
\frac{{\bar\theta}^{2}(^{\bar\alpha}_{\bar\beta})}{{\bar\eta}^2}
\frac{{\bar\theta}^{2}
(^{{\bar\alpha}+\gamma}_{{\bar\beta}+\delta})}{{\bar\eta}^2}
(-)^{\delta(\alpha+{\bar\alpha})}
\frac{\theta^{2}(^{\gamma}_{\delta})}{\eta^{2}}
\frac{{\bar\theta}^{2}(^{\bar\gamma}_{\bar\delta})}{{\bar\eta}^{2}}
\ ,
\label{pft}
\end{eqnarray}
where the $\theta$-functions with arguments $\gamma$ and $\delta$
denote the contribution of the two compactified coordinates
$\phi_{1,2}$ at radius 1.

The partition function (\ref{pft}) can be decomposed as:
\begin{eqnarray}
Z[M^4\otimes T^2\otimes W^{(4)}_{k=0}]=&&
{{\rm Im}\tau^{-3/2}\over \eta^3{\bar\eta}^3}
\sum_{\alpha,{\bar\alpha},\gamma,l,{\bar l}}
(-)^{\alpha+{\bar\alpha}}~
\chi^{l}_{H^{+}_1} \chi^{l+\alpha}_{H^{-}_1}~
\chi^{1+l}_{H^{+}_2} \chi^{1+l+\alpha+\gamma}_{H^{-}_2}~
\nonumber \\
{\bar\chi}^{\bar l}_{H^{+}_1}
{\bar\chi}^{{\bar l}+{\bar\alpha}}_{H^{-}_1}~
{\bar\chi}^{1+{\bar l}}_{H^{+}_2}
{\bar\chi}^{1+{\bar l}+{\bar\alpha}+\gamma}_{H^{-}_2}&&
\sum_{l',{\bar l}'} {1\over 2}
[1+(-)^{\alpha+l+{\bar\alpha}+{\bar l}+l'+{\bar l}'}]~
\chi^{l'}_{\phi^+} \chi^{l'+\gamma}_{\phi^-}~
{\bar\chi}^{{\bar l}'}_{\phi^+}
{\bar\chi}^{{\bar l}'+{\gamma}}_{\phi^-}\ ,
\label{pftchar}
\end{eqnarray}
where $\phi^{\pm}=(\phi_1\pm\phi_2)/{\sqrt 2}$ are both compactified
at the $SU(2)$ self-dual point. The $\delta$-projection is:
\begin{equation}
2{\tilde S}_2 + 2{\bar{\tilde S}}_2 + l'+{\bar l}' = {\rm even}\ .
\label{dprojt}
\end{equation}
The lower-lying states from both the untwisted and twisted sectors
are degenerate with mass-squared equal to $Q^2/8=1/8$. The
heterotic construction can be done using the maps (\ref{pfheta})
and (\ref{pfhetb}).

\subsection{The $F^{(2)}\otimes W^{(4)}_{k_1}\otimes W^{(4)}_{k_2}$
partition function}

The above constructions can be easily extended to the background
solutions $F^{(2)}\otimes W^{(4)}_{k_1}\otimes W^{(4)}_{k_2}$. In
this case, the $Z_2$ projection can act in different ways in the two
spaces consistently with modular invariance. An interesting limit is
when $k_1=k_2=0$, where the Kac-Moody currents decouple. Then, the
$N=4$ algebra is realized only in terms of world-sheet fermions and
the two $U(1)$ currents with background charges $Q_1=Q_2=1$. As
mentioned in Section 2.2, this model can be seen as a non-critical
superstring of ${\hat c}_M=5$ coupled to the Liouville field, which
is
identified with the linear combination of the two $U(1)$'s carrying
the background charge $Q_L={\sqrt{ Q_1^2+Q_2^2}} ={\sqrt 2}$. For
$(k_1+k_2)$ even, the partition function for the $F^{(2)}\otimes
W^{(4)}_{k_1}\otimes W^{(4)}_{k_2}$ model can be derived in a
similar way as in (\ref{pf}) and (\ref{pfkodd}):
\begin{eqnarray}
Z_{W_{k_1} \otimes W_{k_{2}}}= {{\rm Im}\tau^{-1}\over
\eta^2{\bar\eta}^2} {1\over 8}
\sum_{\alpha,\beta,{\bar\alpha},{\bar\beta},\gamma,\delta}
&&(-)^{\alpha+\beta}
\frac{\theta^{2}(^{\alpha}_{\beta})}{\eta^2}
\frac{\theta^{2}(^{\alpha+\gamma}_{\beta+\delta})}{\eta^2}
(-)^{{\bar\alpha}+{\bar\beta}}
\frac{{\bar\theta}^{2}(^{\bar\alpha}_{\bar\beta})}{{\bar\eta}^2}
\frac{{\bar\theta}^{2}
(^{{\bar\alpha}+\gamma}_{{\bar\beta}+\delta})}{{\bar\eta}^2}
\nonumber \\
&&(-)^{\delta(\alpha+{\bar\alpha}+{(k_1+k_2)\over 2}\gamma)}
    Z_{k_1}[^{\gamma}_{\delta}]~Z_{k_2}[^{\gamma}_{\delta}]\ .
\label{pfWW}
\end{eqnarray}
After taking into account the $\beta$ and $\bar{\beta}$
projections, we can express $Z_{W_{k_1} \otimes W_{k_{2}}}$ in terms
of the various $SU(2)$ characters:
\begin{eqnarray}
Z_{W_{k_1} \otimes W_{k_{2}}}&&=
{{\rm Im}\tau^{-1}\over \eta^2{\bar\eta}^2}
\sum_{\alpha,{\bar\alpha},\gamma,l,{\bar l}}
(-)^{\alpha+{\bar\alpha}}~
\chi^{l}_{H^{+}_1} \chi^{l+\alpha}_{H^{-}_1}~
\chi^{1+l}_{H^{+}_2} \chi^{1+l+\alpha+\gamma}_{H^{-}_2}~
{\bar\chi}^{\bar l}_{H^{+}_1}
{\bar\chi}^{{\bar l}+{\bar\alpha}}_{H^{-}_1}~
{\bar\chi}^{1+{\bar l}}_{H^{+}_2}
{\bar\chi}^{1+{\bar l}+{\bar\alpha}+\gamma}_{H^{-}_2}
\nonumber \\
&&{\sum_{L,L'}}{1\over 2}
[1+(-)^{\alpha+l+{\bar\alpha}+{\bar l}+{(k_1+k_2)\over
2}\gamma+L+L'}]~
\chi_{k_1}^L {\bar\chi}_{k_1}^{L+\gamma (k_1-2L)}
\chi_{k_2}^{L'} {\bar\chi}_{k_2}^{L'+\gamma (k_2-2L')}
\label{pfcharWW}
\end{eqnarray}
The lower-lying states in the untwisted sector ($\gamma=0$) have
(left and right) mass-squared equal to $\frac{ Q_1^2+Q_2^2}{8}$.
In the general case, where $k_1$ and $k_2$ are non-zero, the
$\delta$-projection becomes:
\begin{equation}
2{\tilde S}_2 + 2{\bar{\tilde S}}_2 + L+L' + {(k_1+k_2)\over 2}\gamma
= {\rm even}\ ,
\label{dprojWW}
\end{equation}
correlating (for non-vanishing $k_i$) the $SU(2)_{k_i}$ spins
with those of $SU(2)_{H^{-}_i}$. This correlation implies that the
lower lying states in the twisted sector ($\gamma=1$) are heavier
than those of the untwisted sector, provided $k_i$ are large. For
$k_i$ small, and in particular when both $k_i$'s are zero, the
twisted states can have mass-squared lower than
$\frac{Q^{2}_1+Q^{2}_2}{8}$. When $k_1$ and $k_2$ are even, the
(left and right) mass-squared of the lower-lying twisted states
are equal to $(k_1+k_2)/16$, while those of the lower-lying
untwisted states are  equal to
$\frac{1}{4}(\frac{1}{k_1+2}+\frac{1}{k_2+2})$. Thus, for $k_i$
even, the twisted sector is heavier than the untwisted one for any
value $k_i\ne 0$. In the limiting case $k_i=0$, however, the twisted
sector becomes massless as expected from the ${\hat c}_M =5$
non-critical superstring theory \cite{kuseiberg}. When both $k_i$'s
are odd,
the mass-squared of the lower-lying twisted states is
$\frac{1}{16}(\frac{(k_1+1)^2}{k_1+2}+\frac{(k_2+1)^2}{k_2+2})$ and
thus for $k_1=k_2=1$ the twisted and untwisted lower mass states
are degenerate.

As we mentioned in Section 2.2, a ${\hat c}_M =5$ super-Liouville
model exhibiting $N=4$ superconformal symmetry was proposed in
ref.\cite{gervais}. This model has a five-dimensional Lorentz
invariance while the Liouville zero-mode has a discrete spectrum.
In our case, there are four non-compact dimensions but the
Lorentz symmetry is reduced to three-dimensional due to the
background charge in the Liouville coordinate; in addition,
there are two compact dimensions described by four free fermions
forming an $SU(2)\otimes SU(2)$ level-1 group manifold.
It is interesting to notice that the partition function of
the  Bilal-Gervais model, when one of the five dimensions is
compactified at the $SU(2)$ point ($R={\sqrt 2}$), is the same
with our expression (\ref {pfWW}) in the limiting case $k_i=0$.
Despite this similarity, the two models are expected to be
different since the spectra and their symmetries are not the
same in the two models.

Note that when $k_1=k_2=k$ the two $SU(2)_k$'s combine to form an
$SO(4)_k$ group manifold and the large-$k$ limit of (\ref{pfWW})
corresponds to a classical solution of the $SO(4)_k$ gauged
supergravity theory \cite{CSF}, \cite{Roo}.

The heterotic construction can be done as before,
using the maps given in ({\ref {pfheta}}) and ({\ref {pfhetb}}).
For $k_1=k_2=2$, one obtains a different realization of the ${\hat
c}_M =7$ non-critical superstring (type-II or heterotic).

\section{Conclusions}

  String solutions in the semiclassical limit define background
solutions of some special effective field theories. This limit turns
out to be very useful regarding the study of the string-induced
low-energy theories, as well as the study of  physics  in weakly
curved domains of space time. The  field theory picture, however,
completely fails when the involved curvartures are strong; it is then
necessary  to go beyond the semiclassical limit
and work directly on the string level, using the powerful techniques
of the underlying two-dimensional (super)conformal field theory. For
a generic string background the stringy approach is at present
non-accessible, because of some technical difficulties, which
hopefully will be solved in the future. As we show in this work, it
is possible to go further in the stringy direction for some special
choices of the target-space backgrounds, namely when one chooses  the
world-sheet degrees of freedom to form  non-trivial realizations of
the $N=4$  superconformal symmetry.

 In the weak curvature limit, all our solutions have a
ten-dimensional target space interpretation, and each one of them
contains two curved four-dimensional subspaces; each 4-d subspace
defines a ${\hat c}=4$, $N=4$ superconformal system and contains an
integer parameter $k$  defining the strength of its curvature (${\cal
O}({1\over k+2})$). The semiclassical limit is when both
4-d curvatures are small: ($k_1,k_2 \rightarrow \infty$). In our
analysis we used as building blocks four topologically non-trivial
4-d subspaces found in ref.\cite{n4kounnas}: i) The $W_{k}^{(4)}$
space, which has the shape of a 4-d (semi)wormhole; ii) the
$C_{k}^{(4)}$ space, with the shape of (2-d bell)$\otimes$(2-d
cylinder); iii) and iv) the two versions of the $\Delta_{k}^{(4)}$
space, with the shape of
(2-d bell)$\otimes$(2-d cigar) and (2-dbell)$\otimes$(2-d trumpet),
respectively.
In the large-$k_i$ limit, the constructions based on $W_{k_i}^{(4)}$
superconformal blocks describe some stable solutions of 4-d gauged
supergravities, $N=8$ in the type-II construction or $N=4$ in the
heterotic construction.

The constructions based on $W_{k}^{(4)}$ and/or $C_{k}^{(4)}$ blocks
are connected to the non-critical strings and define super-Liouville
theories in the strong coupling regime coupled to  unitary matter
systems. The  Liouville and matter central charges, ${\hat
c_L}=1+4({1\over k_1+2}+{1\over k_2+2})$ and
${\hat c_M}=1-4({1\over k_1+2 }+{1\over k_2+2 })$ are given in terms
of the two-integer parameters $k_1$ and $k_2$.
The lower value ${\hat c}_M=5$ corresponds to the $W_{k_i}^{(4)}$
construction,  in the limiting case where $k_1=k_2=0$.
Another interesting value is when ${\hat c}_M=7$, obtained with
$k_1=0$ and $k_2\rightarrow\infty$, or when $k_1=k_2=2$. It
turns out that this case is in  correspondence with the
high-temperature phase of the heterotic critical superstring
\cite{highT}.

The full spectrum of excitations can be derived in all our
constructions combining unitary representations of the $N=4$
superconformal theory in a modular-invariant way. In the case of
$W_{k}^{(4)}$ constructions, these representations are expressed in
terms of the well-known $SU(2)$ characters, while in all other
constructions one uses also  the characters  of some compact
($SU(2)/U(1)$)  and/or non-compact ($SL(2,R)/U(1)$) parafermions, as
well as those of  free boson compactified in a given special radius.
For the $W_{k}^{(4)}$ constructions, we give the full spectrum of
propagating states in terms of modular-invariant partition functions
for all values of $k_i$'s. When $k_i$'s  are large, all
states in the $Z_2$ twisted sector have mass-squared which
grows with $k_i$. The lower-lying states in this sector have
mass-squared equal to $(k_1+k_2)/16$ and,
thus they decouple in the semiclassical limit. In this limit the
remaining states are those of the $Z_2$ untwisted sector with
masses lower than the string scale; these states are in one-to-one
correspondence with those of a 10-d Kaluza-Klein field  theory
defined in a double wormhole space; their masses are just given in
terms of the two $SU(2)_{k_i}$  spherical excitations with
a shifting due to the non trivial dilaton and/or torsion
background:  $M^{2}_{l_1,l_2}={1\over 4}\left({(l_1+1)^2\over  k_1+2
}+{(l_2+1)^2\over  k_2+2 }\right)$.  The untwisted states are then
similar to the quantized momenta of a toroidal
compactification while the twisted states are in correspondence
with the string-winding modes.

For small $k_i$, the lower-mass of the states,
in both  the twisted and
untwisted sector, is of the same order as the string mass scale and
thus, the wormhole target space-time interpretation fails.
The untwisted states can  be massless only for $k_i=\infty$.
The twisted states are in general massive.
We  found, however, two special cases were
some of the twisted states are massless. The existence of the first
case  was conjectured in the framework of super-Liouville theories in
ref.\cite{kuseiberg} and it corresponds to the limiting case where
both $k_i$ are zero, with ${\hat c_M}= 5$. The second special case is
new and it corresponds to a ${\hat c_M}= 8$ super-Liouville theory
with $k_1=2$ and $k_2=\infty$.\\[1cm]
{\bf Acknowledgements}

Research supported in part by  the EEC contracts SC1--CT92--0792,
SC1$^{*}$-0394C,
SC1$^{*}$-CT92-0789, and in part by the DOE grant
DOE-AT03-88ER40384, taskC.

I.A. and C.K. thank the CERN Theory Division and the Centre de
Physique
Th\'eorique at the Ecole Polytechnique, respectively, for their
hospitality during completion of this work. We would like to thank C.
Callan, D. Gross, E. Kiritsis and D. Lust for fruitful discussions.


\newpage
\begin{thebibliography}{99}

\bibitem{betafunct} E. Fradkin and A. Tseytlin,
\np {\bf 261} (1985) 1;\\
C. Callan, D. Friedan, E. Martinec and M. Perry,
\np {\bf 262} (1985) 593.

\bibitem{alpaprim} R.C. Myers, \pl {\bf 199} (1987) 371;\\
 M. Muller, \np {\bf 337} (1990) 37;\\
I. Bars, \np {\bf 334} (1990) 125;\\
K.A. Meissner and G. Veneziano, \pl {\bf 267} (1991) 33;\\
A. Sen, \pl{\bf 271} (1991) 295;\\
M. Gasperini, J. Maharana and G. Veneziano, \pl {\bf 272} (1991)
277;\\
M. Gasperini and G. Veneziano, \pl {\bf 277} (1992) 256;\\
A.A. Tseytlin, Class. Quant. Grav. {\bf 9} (1992) 979;
Int. J. Mod. Phys. {\bf D1} (1992) 223; Cambridge preprint
DAMPT-92-36;\\
K. Behrndt, \pl{\bf 301} (1993) 29; DESY preprint DESY
92-055;\\
M. Gasperini and G. Veneziano, {\it Astropart. Phys.} {\bf 1} (1993)
317;\\
A.A. Tseytlin, Cambridge preprints DAMPT-92-15 and DAMPT-92-36;\\
M. Gasperini and G. Veneziano, preprint CERN-TH.6572/92;\\
H.J. de Vega, A.V. Mikhailov and N. Sanchez, Univ. Paris VI preprint
LPTHE
92-32;\\
H.J. de Vega and N. Sanchez, \pr {\bf 47} (1993) 3394;\\
H.J. de Vega, A.V. Mikhailov and N. Sanchez, Univ. Paris VI preprint
LPTHE 92-32.

\bibitem{KKLn24} E. Kiritsis, C. Kounnas and D. L\"ust, CERN
preprint,
hep-th/9308124, to appear in \IJMP

\bibitem{ABEN} I. Antoniadis, C. Bachas, J. Ellis and D.V.
Nanopoulos,
\pl {\bf 211} (1988) 393, \np{\bf 328} (1989) 117;\\
J. Polchinski, \np{\bf 324} (1989) 123.

\bibitem{ABS} I. Antoniadis, C. Bachas and A. Sagnotti,
\pl {\bf 235} (1990) 255;\\
C. Bachas and E. Kiritsis, preprint hep-th/9311185.

\bibitem{wormclas} C. Callan, J. Harvey and A. Strominger,
\np {\bf 359} (1991) 611;\\
C. Callan, Lectures at Sixth J.A. Swieca Summer School, Princeton
preprint PUPT-(1991).
\bibitem{blackhole} E. Witten, \pr {\bf 44} (1991) 314.

\bibitem{coset} I. Bars, \np {\bf 334} (1990) 125;\\
I. Bars and D. Nemeschansky, \np {\bf 348} (1991) 89.

\bibitem{blackholever} R. Dijkgraaf, E. Verlinde and H. Verlinde,
\np {\bf 371} (1992) 269.

\bibitem{n4kounnas} C. Kounnas, \pl {\bf 321} (1994) 26;\\
Proceedings of the  International Conference, on ``Lepton-Photon
Symposium and Europhysics Conference on High Energy
Physics", Geneva, 1991, Vol. 1, pp. 302-306; Proceedings of the
International Workshop on "String Theory, Quantum Gravity and
Unification of Fundamental Interactions", Rome, 21-26 September 1992,
CERN preprint TH.6790/93.

\bibitem{Reydilatoninst} S.J. Rey, \PR {\bf D43} (1991) 526 and
SLAC-PUB-5659 (1991).

\bibitem{freaxioninst} M. Bill\'o, P. Fr\'e, L. Girardello and A.
Zaffaroni,
Int. J. Mod. Phys. {\bf A8} (1993) 2351.

\bibitem{ginscoset} P. Ginsparg and F. Quevedo, \np {\bf 385}
(1992) 527.
\bibitem{tsva} R. Brandenberger and C. Vafa, \np {\bf 316} (1988)
391;\\
A.A. Tseytlin and C. Vafa, \np {\bf 372} (1992) 443.
\bibitem{koulu} C. Kounnas and D. L\"ust, \pl {\bf 289} (1992) 56.\\
D. L\"ust, Proccedings of ``4th
Hellenic School on Elementary Particle Physics", Corfu,
1992, CERN preprint TH.6850/93.

\bibitem {nawit}C. Nappi and E. Witten, \pl {\bf 293} (1992) 309.

\bibitem {mul} M. Muller, \np {\bf 337} (1990) 37.

\bibitem{hor} P. Horava, \pl {\bf 278} (1992) 101.

\bibitem{nwgrwaves} C. Nappi and E. Witten, IAS preprint,
hep-th/9310112;\\
E. Kiritsis and C. Kounnas, \pl {\bf 320} (1994) 264;\\
K. Sfetsos, Utrecht preprint, hep-th/9311010;\\
C. Klimcik and A. Tseytlin, CERN preprint, CERN-TH.7069/93.

\bibitem{ademolo} M. Ademollo et al., \np {\bf 114} (1976) 297;\\
T. Eguchi and A. Taormina, \pl {\bf 200} (1988) 634;\\
A. Sevrin, W. Troost and A. Van Proeyen, \pl {\bf 208} (1988) 447.

\bibitem{5brane} M.J. Duff and J.X. Lu,
\np {\bf 354} (1991) 129 and 141.

\bibitem{kprn4} C. Kounnas, M. Porrati and B. Rostand, \pl {\bf 258}
(1991) 61.

\bibitem{RoVe} M. Rocek, K. Schoutens and A. Sevrin, \pl {\bf 265}
(1991) 303;\\
M. Rocek and E. Verlinde, \np {\bf 373} (1992) 630.

\bibitem{CSF} E. Cremmer, S. Ferrara and J. Scherk,
\pl {\bf 74} (1978) 61;\\
D. Freedman and J. Schwarz \np {\bf 137} (1978) 333;\\
B. de Wit and H. Nicolai \np {\bf 208} (1982) 323.

\bibitem{Roo} M. de Roo, \np {\bf 255} (1985) 515;
\pl {\bf 156} (1985) 331;\\
E. Bergshoeff, I.G. Koh and E. Sezgin, \pl {\bf 155} (1985) 71.

\bibitem{noncrit1} A. M. Polyakov, \pl {\bf 103} (1981) 207;
{\it Mod. Phys. lett.} {\bf A2} (1987) 893;\\
V.G. Knizhnik,  A. M. Polyakov and A.B. Zamolodchikov, {\it Mod.
Phys. lett.} {\bf A3} (1988) 819;\\
J.L. Gervais and A. Neveu, \np {\bf 264} (1986) 577; \np {\bf
257} [FS14] (1985) 59.

\bibitem{gervais} A. Bilal and J.L. Gervais, \pl {\bf 177} (1986) 313;
\pl {\bf 187} (1987) 39; \np (1987) {\bf 293} 1.

\bibitem{noncrit2} F. David,
{\it Mod. Phys. Lett.} {\bf A3} (1988) 1651;\\
J. Distler and H. Kawai, \np {\bf 321} (1989) 509.

\bibitem{2dgrav} E. Brezin and V. Kazakov, \pl {\bf 236} (1990) 144;\\
M. Douglas and S. Shenker, \np {\bf 335} (1990) 635;\\
D. Gross and A. Migdal, {\it Phys. Rev. Lett.} {\bf 64} (1990) 127.

\bibitem{kuseiberg} D. Kutasov and N. Seiberg,
\pl {\bf 251} (1990) 67.

\bibitem{highT} I. Antoniadis and C. Kounnas,
\pl {\bf 261} (1991) 369.

\bibitem{dualTW} J.J. Atick and Witten, \np{\bf 310} (1988)291;\\
C. Kounnas and B. Rostand \np {\bf 341} (1990) 641.

\bibitem{zamoParafemions} A.B. Zamolodchikov and V.A. Fateev,
\SPJ~ {\bf 62}(1985)215; \SPJ ~{\bf 63}(1986) 913;\\
P. Di Vecchia, J. Petersen, M. Yu and H. Zheng, \pl {\bf 174} (1986)
280;\\
Z. Qiu, \pl {\bf 188} (1987) 207.\\
V. Kac and D. Peterson,  \AM {\bf 53} (1984) 125.

\bibitem{DLPnoncompact} L. Dixon, J. Lykken and M. Peskin, \np {\bf
325} (1989) 329;\\
J. Lykken, \np {\bf 313} (1989) 473;\\
I. Bakas and E. Kiritsis,  \IJMP {\bf A7} [Suppl.1A] (1992) 339;
\pl {\bf 301} (1993) 49.

\bibitem{giveondual} A. Giveon, \MPL {\bf A6} (1991) 2843;\\
E. Kiritsis, \MPL {\bf A6} (1991) 2871;
Nucl. Phys. {\bf B405} (1993) 109;\\
A. Giveon and E. Kiritsis, \np {\bf 411} (1994) 487.

\bibitem{banksdixon} T. Banks and L. Dixon, \np {\bf 307} (1988) 93.

\bibitem{gepner4d} D. Gepner, \pl {\bf 199} (1987) 370;
\np {\bf 296} (1988) 757.

\bibitem{orbifold} L.J. Dixon, J. Harvey, C. Vafa and E. Witten, \np
{\bf 261} (1985) 678; {\bf 274} (1986) 285.

\bibitem{abk4d} I. Antoniadis, C. Bachas and C. Kounnas, \np {\bf
289}
(1987) 87;
\\ H. Kawai, D.C. Lewellen and S.H.-H. Tye, \np {\bf 288} (1987) 1.

\bibitem{llsmap} W. Lerche, D. L\"ust and A.N. Schellekens, \pl {\bf
181} (1986) 71; \np {\bf 287} (1987) 477.


\end{thebibliography}
\end{document}